\documentclass[longauth]{aa}
\usepackage[colorlinks=true,linkcolor=blue,citecolor=blue]{hyperref}
\usepackage{lipsum,multicol}
\usepackage{graphicx}

\usepackage{natbib}
\def \gs{g\,s$^{-1}$}
\def \kms{km\,s$^{-1}$}
\def \ms{m\,s$^{-1}$}
\def \cms{cm\,s$^{-2}$}
\def \ergcms{erg\,s$^{-1}$ cm$^{-2}$}

\def \mjup{M$_\mathrm{jup}$}
\def \rjup{R$_\mathrm{jup}$}
\def \Hei{\ion{He}{I}$\,\,$}
\def\sun{\odot}
\bibpunct{(}{)}{;}{a}{}{,} 
\usepackage{txfonts}
\usepackage{soul}
\usepackage{pdflscape}
\usepackage{capt-of}
\usepackage{placeins}
\DeclareMathAlphabet{\mathpzc}{OT1}{pzc}{m}{it}
\begin{document} 


   \title{The GAPS programme at TNG \thanks{Based on observations made with the Italian Telescopio Nazionale
{\it Galileo} (TNG) operated by the {Fundaci\'on Galileo Galilei} (FGG) of the
{Istituto Nazionale di Astrofisica} (INAF) at the
{Observatorio del Roque de los Muchachos} (La Palma, Canary Islands, Spain).}}

   \subtitle{LXIII. Photo-evaporating puzzle: Exploring the enigmatic nature of TOI-5398 b's atmospheric signal}
   \author{M. C. D'Arpa
          \inst{1,}\inst{2},
          G. Guilluy\inst{3},
          G. Mantovan\inst{4,5},
          F. Biassoni\inst{6,7},
          R. Spinelli\inst{1},
          D. Sicilia\inst{8},
          D. Locci \inst{1},
          A. Maggio \inst{1},
          A.F. Lanza \inst{8},
          A. Petralia \inst{1},
          C. Di Maio \inst{1},
          S. Benatti \inst{1},
          A.S. Bonomo \inst{3},
          F. Borsa \inst{7},
          L. Cabona \inst{5},
          S. Desidera \inst{5},
          L. Fossati \inst{9},
          G. Micela \inst{1},
          L. Malavolta \inst{4},
          L. Mancini \inst{3,10,11},
          G. Scandariato \inst{8},
          A. Sozzetti \inst{3},
          M. Stangret \inst{5},  
          L. Affer \inst{1},
          F. Amadori \inst{2,5},
          M. Basilicata \inst{10,5},
          A. Bignamini  \inst{12},
          W. Boschin   \inst{13},
          A. Ghedina \inst{13}
          }
          
   \institute{INAF -- Palermo Astronomical Observatory, Piazza del Parlamento, 1, 90134 Palermo, Italy\\ 
              \email{mattia.darpa@inaf.it}
         \and
              University of Palermo, Department of Physics and Chemistry “Emilio Segrè” 
        \and
            INAF -- Osservatorio Astrofisico di Torino, Via Osservatorio 20, 10025, Pino Torinese, Italy 
        \and
            Dipartimento di Fisica e Astronomia ``Galileo Galilei'', Università di Padova, Vicolo 
            dell'Osservatorio 3, IT-35122, Padova, Italy 
        \and
            INAF -- Osservatorio Astronomico di Padova, Vicolo dell'Osservatorio 5, 35122, Padova, Italy 
        \and
            DISAT, Università degli Studi dell'Insubria, via Valleggio 11, 22100, Como, Italy 
        \and
            INAF -- Osservatorio Astronomico di Brera, Via E. Bianchi 46, 23807 Merate (LC), Italy 
        \and
            INAF -- Osservatorio Astrofisico di Catania, Via S. Sofia 78, I-95123, Catania, Italy 
        \and 
            Space Research Institute, Austrian Academy of Sciences, Schmiedlstrasse 6, 8042 Graz, Austria 
        \and
            Department of Physics, University of Rome “Tor Vergata”, Via della Ricerca Scientifica 1, 00133 Rome, Italy 
        \and
            Max Planck Institute for Astronomy, Königstuhl 17, 69117 Heidelberg, Germany 
        \and
            NAF – Osservatorio Astronomico di Trieste, via Tiepolo 11, 34143 Trieste 
        \and
            Fundación Galileo Galilei-INAF, Rambla José Ana Fernandez Pérez 7, 38712 Breña Baja, TF, Spain 
     }

   \date{XXXX; XXXX}

 
  \abstract
   {Atmospheric characterisation plays a key role in the study of exoplanetary systems, giving hints about the current and past conditions of the planets. The information retrieved from the analysis of pivotal lines such as the H$\alpha$ and \Hei triplet allow us to constrain the evolutionary path of the planets due to atmospheric photo-evaporation. After focussing for many years on ultra-hot Jupiters, atmospheric characterisation is slowly moving towards smaller and colder planets, which are harder to study due to the difficulties in extracting the planetary signal and which require more precise analysis.}
   {We aim to characterise the atmosphere of TOI-5398 b (P $\sim$ 10.59 days), the outer warm Saturn orbiting a young ($\sim$ 650 Myr) G-type star that also hosts the small inner planet TOI-5398 c (P $\sim$ 4.77 days). Both planets are suitable for atmospheric probing due to the closeness to their host star, which results in strong photo-evaporation processes, especially the larger outer one with an estimated transmission spectroscopy metric of 288 (higher than those of several well-known hot Jupiters).}
   {We investigated the atmosphere of planet b, analysing the data collected during a transit with HARPS-N and GIANO-B high-resolution spectrographs, employing both cross-correlation and single-line analysis to study the presence of atomic species. Incidentally, we recorded the simultaneous transit of planet c, and hence we also focussed on discerning the origin of the signal. We expect planet b to be the cause of the detected signal, since, according to existing evaporation models, it is currently expected to lose more mass than planet c. }
   {We detected the presence of H$\alpha$ and \Hei triplets, two markers of the photo-evaporation processes predicted for the system, retrieving a height in the atmosphere of 2.33 Rp and 1.65 Rp, respectively. We confirmed these predictions by employing the models computed with the ATES software, which predict a He I absorption arising from planet b comparable with the observed one. Moreover, the ATES models suggested an He/H ratio of 1/99 to match our observations. The investigation of atomic species led to the detection of an Na I doublet via single-line analysis, while the cross-correlation did not return a detection for any of the atomic species investigated.}
   {}


   \titlerunning{The enigmatic nature of TOI-5398 b atmospheric signal}
    \authorrunning{M.C.D'Arpa et al.}

   \maketitle
%


\section{Introduction}
The characterisation of exoplanetary atmospheres is a key frontier in astrophysics. It offers valuable insights into the physical and chemical processes occurring in distant worlds. The field has witnessed significant advancements, with various observational techniques contributing to our understanding of their diverse compositions. Among these, high-resolution spectroscopy emerges as a powerful tool, since this technique allows one to resolve individual spectral lines, providing detailed information about the chemical composition and atmospheric dynamics. The ability to break degeneracies induced by broad spectral features in low-resolution spectra makes high-resolution spectroscopy indispensable for accurate atmospheric studies, which focus on the analysis of individual atomic lines and matching with model templates.
High-resolution transmission spectroscopy has predominantly been applied to hot and massive exoplanets, such as hot Jupiters (HJs), because of their high equilibrium temperatures due to the small distances from their host star \citep{Guilluy2022}. The closeness and the consequently high equilibrium temperature, $T_{\rm eq}$, favour photo-evaporation processes, and enhance the scale height defined as
\begin{equation}
    H = \frac{k_{\rm B}T_{\rm eq}}{\mu g_{\rm p}},
\end{equation}
where $k_B$ is the Boltzmann's constant and $\mu$ is the mean molecular weight. In many cases, this leads to a large annulus that scatters the starlight during the transit, resulting in a strong signal. However, the large HJ's mass increases the surface gravity, $g_{\rm p}$, and hence diminishes the scale height. Therefore, the ideal targets for transmission spectroscopy are close and hot inflated planets with lower densities. The poorly explored class of warm, Saturn-like exoplanets combines these characteristics, presenting an exciting opportunity to broaden our understanding of planetary atmospheres beyond the better-studied HJs.

\begin{figure}
\centering
\includegraphics[width=\linewidth]{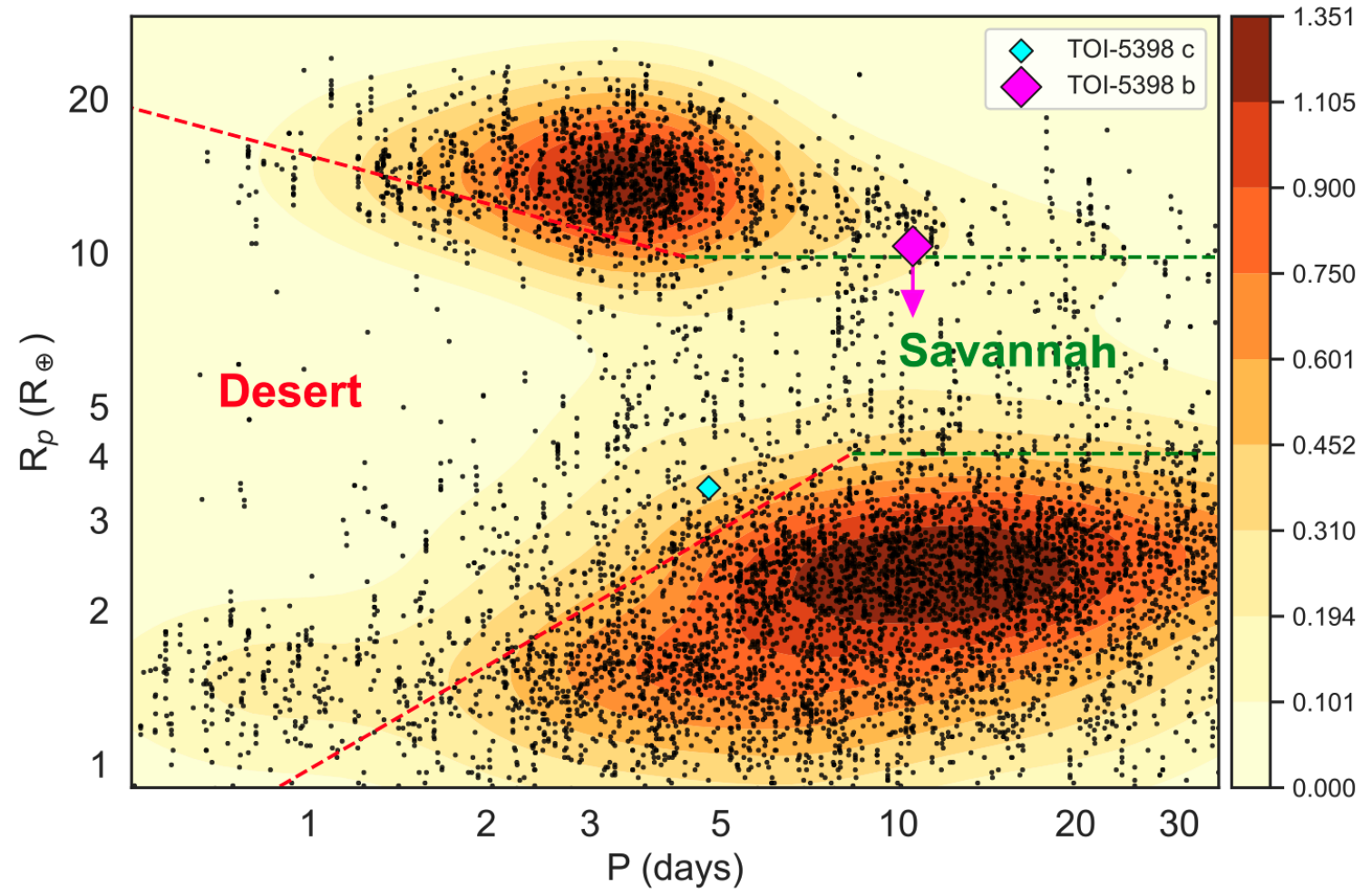}
\caption{Known exoplanets as a function of their radius and period from the NASA Exoplanet Archive \citep{Akeson2013} within the Neptunian desert and savannah. The color represents the density of planets. TOI-5398 b is highlighted with a purple diamond, while planet c is highlighted with a cyan one. The arrow indicates the expected position of planet b within the savannah at the end of its evolution according to \cite{mantovan_ross}. The marker size is proportional to the planetary mass.}
\label{fig_desert}
\end{figure}

The TOI-5398 system (see Table \ref{tab_par} for the parameters) is a fascinating target for atmospheric studies due to its compact planetary architecture consisting of two planets orbiting close to the host star \citep{mantovan24}, with the outer one 0.0980 au distance from the star. The outer planet, TOI-5398 b, is a warm, Saturn-like exoplanet with distinct features that make it an ideal candidate for transmission spectroscopy. It has an equilibrium temperature of 947 K, resulting in a scale height of $H_b$ = 1106 km (with $\mu$ = 1.3 assuming a hydrogen-dominated atmosphere, instead of a hydrogen-and-helium-dominated atmosphere as in \citealt{Fossati2022}), which is larger than the mean value of an HJ ($\sim$ 450 km) and the largest transmission spectroscopy metric (TSM, \citealt{2018PASP..130k4401K}) value (TSM = 288) among all known warm giant planets. The target's position at the border of the Neptunian valley and savannah \citep{2017ApJ...847...29O, 2019ApJ...880L...1A} makes it even more fascinating, since its predicted evolution will lead it to the centre of the savannah in the coming hundreds of millions of years, as is shown in Fig. \ref{fig_desert} and in Fig (E.1) from \cite{mantovan_ross}.
The system evolution was strongly shaped by the closeness to the host star, a young (650 Myr) G-type star responsible for the photo-evaporation of the planets, as is discussed by \cite{mantovan_ross}. Based on their findings, planet b underwent minimal mass loss, and hence the present size and composition of this gas giant are probably primordial, while planet c, on the other hand, was more profoundly influenced by photo-evaporation, as is hinted at by its modal density.

The stellar X-ray and extreme ultraviolet (XUV) flux plays a crucial role in influencing the atmospheric properties of young exoplanets. Understanding the interactions between the XUV flux and the atmospheres of distant worlds provides critical insights into the atmospheric escape processes, photochemistry, and overall atmospheric evolution. The present study aims to explore the atmospheric composition of this intriguing exoplanet and contribute to the growing body of knowledge on diverse planetary systems. We aim to explore the composition of the upper layers of the atmospheres using different techniques involving both the visible and near-infrared (nIR) parts of our spectra, deepening our knowledge of the role of the XUV flux in the TOI-5398 system. For this purpose, we focus on lines arising from atomic species that are sensitive to photo-evaporation such as \Hei and H$\alpha$, and also search for other atomic species.

The paper is organised as follows. In Sect. \ref{sec:observations}, we describe the observational data. In Sect. \ref{sec:methods}, we illustrate the different methodologies employed to analyse the data in both the visible and nIR parts of the spectrum. In Sect. \ref{sec:results}, we report the results we obtained via the single-line analysis and cross-correlation technique, while in Sect. \ref{sec:discussion} we evaluate the match between our results and the predicted photo-evaporation models. Eventually, we give a short summary that highlights the main conclusions in Sect. \ref{sec:conclusions}.

\begin{table*}[h]
        \centering
 \caption{Stellar and planetary parameters adopted.}
        \begin{tabular}{l c c c }
                \hline \hline
  Parameters & & & Reference\\
  \hline
        \textbf{Stellar Parameters} &  & &\\
        Stellar mass M$_\star$ (M$_\sun$)\dotfill &  1.146 $\pm$ 0.013& & \citet{mantovan24}\\
        Stellar radius, R$_\star$ (R$_\sun$)\dotfill & 1.051 $\pm$ 0.013 & &\citet{mantovan24}\\
        Effective temperature, Teff (K)\dotfill & 6000 $\pm$ 75&  & \citet{mantovan24}\\
        Metallicity, [Fe/H](dex) \dotfill  &0.09 $\pm$ 0.06 & & \citet{mantovan24}\\
        log~g (log$_{10}$(\cms))\dotfill & 4.44 $\pm$ 0.10&  & \citet{mantovan24}\\
        Systemic velocity, v$_\mathrm{sys}$ (\kms )\dotfill & 9.95$\pm$0.30 & & GaiaDR3\\
 $v$ \rm{sin} $i_*$ (\kms ) \dotfill & 7.5 $\pm$ 0.6 & & \citet{mantovan24}\\
        \textbf{Planetary Parameters} & & \\
        & \underline{\textit{TOI-5398 b}} & \underline{\textit{TOI-5398c}} & \\
        \\
        
        Orbital period, P (days)\dotfill & 10.590547$^{+0.000012}_{-0.000011}$ & 4.77271$^{+0.000016}_{-0.000014}$ &\citet{mantovan24}\\
        Transit epoch, T$_0$ (BJD$_\mathrm{TDB}$) \dotfill& 2459616.49232$^{+0.00022}_{-0.00021}$ & & \citet{mantovan24}\\
        Eccentricity, e \dotfill&   $\leq$0.13 & $\leq$0.14 &\citet{mantovan24}\\
        Argument of periastron, $\omega_\star$ (deg) \dotfill&  92.0$^{+82.0}_{-45.0}$ & 172$^{+79}_{-107}$ &\citet{mantovan24}\\
        Stellar reflex velocity, K$_\star$ (\ms) \dotfill& 15.7$\pm$1.5 & 4.1$^{+1.7}_{-1.6}$& \citet{mantovan24}\\
        Orbital major semi-axis, a (au)\dotfill & 0.0980$\pm$0.0050 & 0.057 $\pm$0.003 & \citet{mantovan24}\\
        Orbital inclination, $i$ (deg) \dotfill& 89.21$^{+0.31}_{-0.21}$ & $\geq$88.4  &\citet{mantovan24}\\
        Planetary mass, M$_\mathrm{p}$ (\mjup) \dotfill& 0.185$\pm$0.018 &  0.0345 $\pm$ 0.0125 &\citet{mantovan24} \\
        Planetary radius, R$_\mathrm{p}$ (\rjup) \dotfill& 0.9189$\pm$0.0357 & 0.32$\pm$0.02 &\citet{mantovan24}\\
        Planetary density, $\mathrm{\rho_{p}}$ ($\mathrm{g/cm^3}$) \dotfill& 0.29$\pm$0.05 & 1.50$\pm$0.68
        &\citet{mantovan24}\\
        Impact parameter, b \dotfill&  0.272$^{+0.069}_{-0.110}$ & $\leq$0.34  &\citet{mantovan24}\\
        Projected obliquity, $\lambda$ (deg) \dotfill&  3.0$^{+6.8}_{-4.2}$ &   &\citet{mantovan_ross}\\
        Equilibrium temperature, T$_\mathrm{eq}$ (K) \dotfill& 947$\pm$28 & 1242 $\pm$ 37&\citet{mantovan24}\\ 
        Planet radial-velocity semi-amplitude, K$_\mathrm{p}$(\kms) \dotfill& 101.58 & 132.14 & This paper \tablefootmark{a} \\
        \hline         \\
                \hline
        \end{tabular}
        \tablefoot{\tablefoottext{a}{
K$_\mathrm{p}=\frac{2\pi a}{P}\frac{\sin{i}}{\sqrt{1-e^2}}=(\frac{2\pi G}{P})^{\frac{1}{3}}\frac{(M_\star+M_\mathrm{p})^{\frac{1}{3}}\sin{i}}{\sqrt{1-e^2}}$.}
}
        \label{tab_par}
\end{table*}


\section{Observations} \label{sec:observations}
We observed the system TOI-5398 as part of the DDT proposal A46DDT4 (PI: G. Mantovan) on March 25 2023 UT with the GIARPS observing mode \citep{GIARPS_claudi} of the Telescopio Nazionale Galileo (TNG). GIARPS allows for simultaneous monitoring at high resolution in the visible band (0.39-0.69~$\mathrm{\mu}$m) with HARPS-N ($R \approx 115,000$) and in the nIR (0.95-2.45$\mathrm{\mu}$m) with GIANO-B ($R \approx 50,000$).

For the GIANO-B observations, we employed an ABAB nodding pattern \citep{GIARPS_claudi} to optimise thermal background noise subtraction. 
GIANO-B is an echelle spectrograph that covers four spectral bands in the nIR (Y, J, H, and K) divided into 50 orders. For our analysis, we focussed on order \#39 in the Y band, where the \Hei triplet is located. The HARPS-N spectrum, on the other hand, is made up of 69 orders along its wavelength range. We focussed on orders \#56 and \#64 for the single-line analysis of the Na {\sc i} doublet and H$\alpha$, respectively, while we employed the whole spectrum for the cross-correlation with template analysis. 
TOI-5398 b was scheduled for observations before, during, and after the planetary transit. Coincidentally, during the selected transit window, TOI-5398 c was also transiting (see Fig \ref{fig:obs}). The impact of planet c's transit on the analysis of planet b is further discussed in the next sections.

We observed the target with an exposure time of 600 s for HARPS-N and 300 s for GIANO-B, collecting 47 and 82 spectra, with an average signal-to-noise ratio (S/N) of 43 and 33 for HARPS-N and GIANO-B, respectively. A summary log of the observations is provided in Table~\ref{log}. The same dataset was used in \cite{mantovan_ross} to analyse the geometry of the system and evaluate the impact of photo-evaporation on the two planets of the system.

 \begin{table}
 \caption{Observations log.}
                        \small
        \resizebox{\linewidth}{!}{
                        \begin{tabular}{ c  | c  |  c }
                                \hline \hline 
                                \centering N$_{\mathrm{obs}}$ (In/Out of transit)  & Exp time [s]& S/N$_{\mathrm{avg}}$\\                
                        HARPS-N  GIANO-B & HARPS-N  GIANO-B & HARPS-N  GIANO-B  \\
                                \hline                  
                                  47(24/23) \hspace{1cm} 82(46/36) & 600 \hspace{1cm} 300 &  43  \hspace{1cm} 33 \\
                                \hline
                                
                        \end{tabular}
        }
        \tablefoot{From left to right: the number of observed spectra
                                (N$_{\mathrm{obs}}$), the exposure time, and the average S/N (S/N$_{\mathrm{avg}}$) across the selected spectral ranges (548.1-554.2\,nm and 1082.49-1085.5\,nm for HARPS-N and GIANO-B, respectively).}
        \label{log}
\end{table}

\begin{figure}
\label{fig:sn}
    \includegraphics[width=\linewidth]{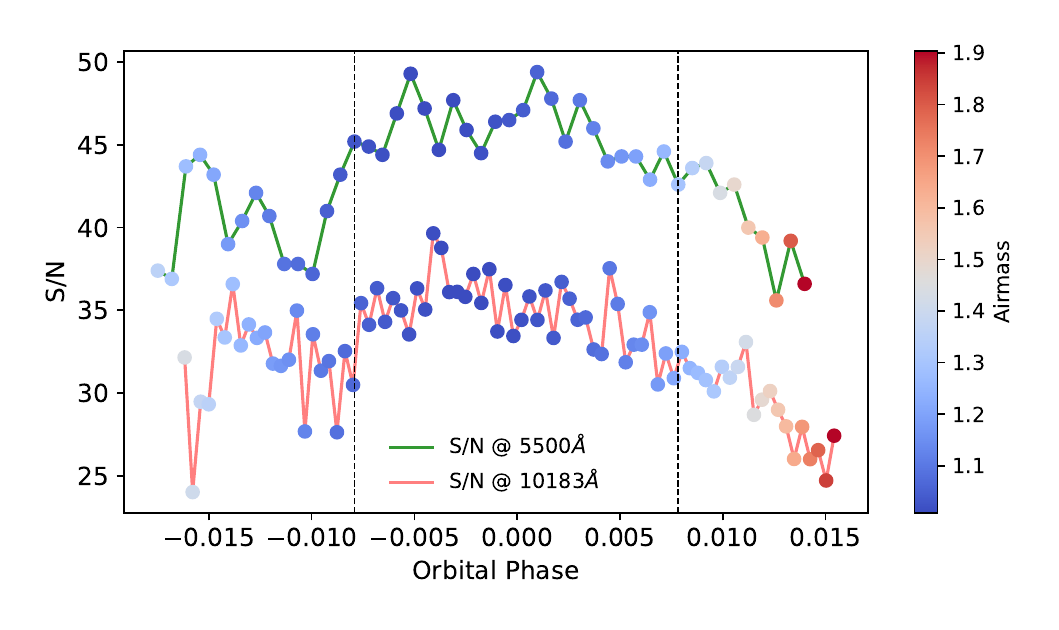}
    \caption{S/N as a function of the orbital phase for both the HARPS-N and GIANO-B datasets. The air mass is colour-coded. The dashed lines represent the ingress and egress of TOI-5398 b.}
\end{figure}

\begin{figure}
\label{fig:obs}
    \includegraphics[width=\linewidth]{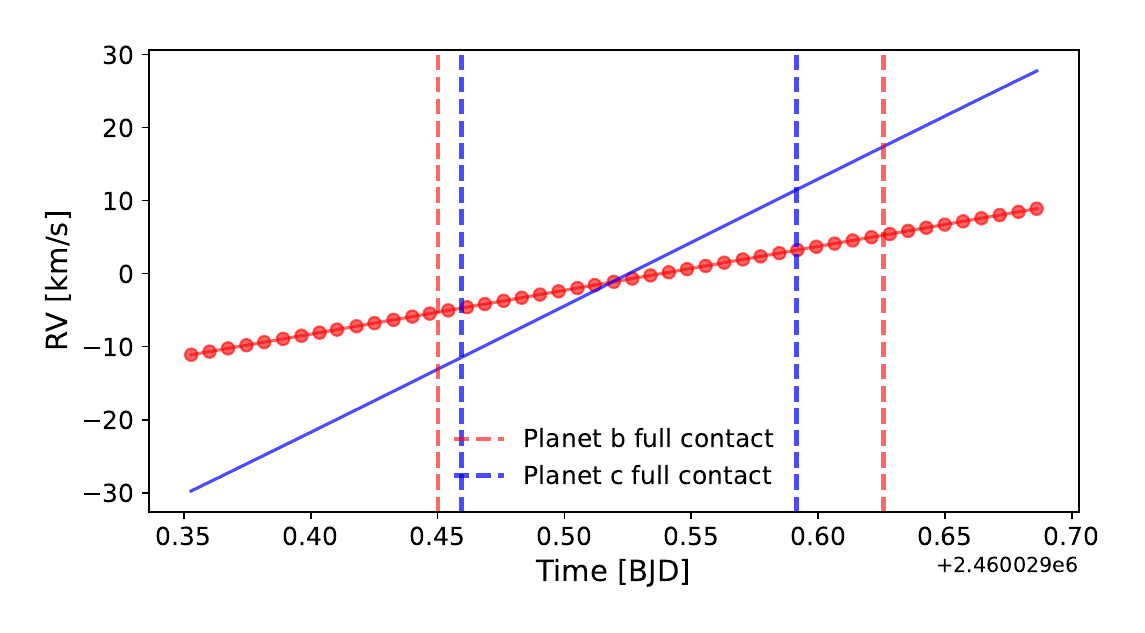}
    \caption{Radial velocity profile of the two planets observed as part of the DDT proposal A46DDT4. The transit of planet c completely overlaps with planet b one.}
\end{figure}

\section{Methods} \label{sec:methods}

\subsection{Extraction of single-line transmission spectra}
The GIARPS observing mode of the TNG is crucial for simultaneously observing both the visible and nIR parts of the planetary spectrum. Recent studies by \cite{Guilluy2024} have demonstrated that combining data from HARPS-N and GIANO-B enables the retrieval of valuable information about the atmospheres of exoplanetary targets. In particular, the joint analysis of the \Hei triplet and H$\alpha$ plays an important role in understanding if a detection may be affected by stellar activity \citep[e.g.,][]{Guilluy2020,Guilluy2023}. The two different datasets require different frameworks to extract the planetary signal around the studied lines. We discuss the different methodologies to extract the transmission spectra in the next subsections. Eventually, we apply a uniform analysis on the extracted spectra. 

We have not included in our analyses any correction to take into account the presence of planet c. Despite planet c being closer, and hence hotter than planet b, the inner planet is almost five times denser than its outer companion, resulting in a smaller scale height ($\sim$ 880 km vs $\sim$ 1100 km). Therefore, we think that planet c does not play a major role in our analysis, as we deeply discuss in Sect. \ref{sec:discussion}.

The main difference between the nIR and visible datasets concerns the modelling and removal of the Rossiter-McLaughlin effect (RME) and the centre-to-limb variation (CLV).
In fact, we neglect the impact of the RME and CLV on the Helium triplet, as several studies have shown that these effects have little (or no) impact on the helium lines \citep[e.g.,][]{Nortmann2018, salz2018, Allart2019, Fossati2022, Allart2023, Guilluy2023}
As was stated before, we do not correct for planet c RME+CLV in the visible dataset as the RME+CLV impact depends on the size of the obscured stellar disc and planet c has a radius three times smaller than planet b. Therefore, the surface covered by planet c is almost ten times smaller than the one obscured by planet b. This means that the intensity of the RME effect should be ten times smaller than planet b's one. Moreover, \cite{mantovan_ross} discussed the same problem regarding their Rossiter analysis to retrieve the orbital obliquity and showed that the radial velocity RME of planet c is comparable with the stellar activity.

\subsubsection{GIANO-B nIR region}
\begin{figure}
    \includegraphics[width=\linewidth]{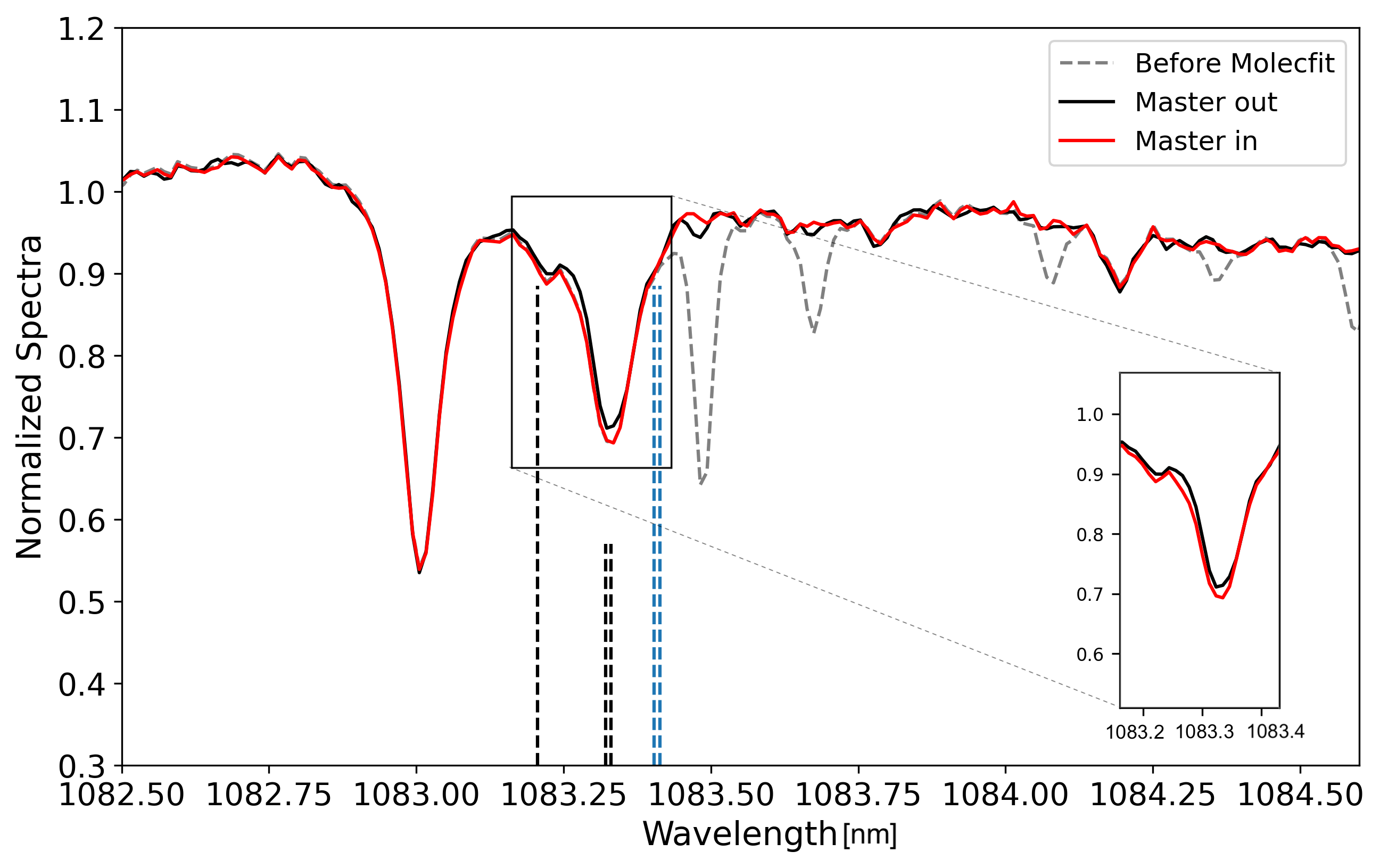}
    \caption{Master-out (black) and master-in (red) spectra in the star frame. The dashed grey line indicates the spectrum before the Molecfit correction, while vertical dashed black and blue lines denote the position of the stellar \ion{He}{I} triplet and OH emission line, respectively. In-transit absorption is visible by eye, especially in the zoomed-in panel. }
    \label{real_spectra}
\end{figure}
We employed the {\tt\string GOFIO} pipeline \citep{Rainer2018} to extract spectra from the raw GIANO-B images. This process involved dark subtraction, flat-field correction, and the removal of bad pixels. Additionally, we performed a preliminary wavelength calibration using a U-Ne lamp spectrum as a template in the vacuum wavelength frame. We focussed on the ms1d spectra, with the echelle orders separated. To enhance the initial wavelength solution, we employed the same approach described in our previous works \citep[e.g.,][]{Giacobbe2021, Guilluy2022}. We aligned all the spectra to the Earth's atmospheric rest frame, assuming it as the frame of the observer (disregarding any $\sim$10 m s$^{-1}$ differences due to winds), by measuring any shifts relative to an average spectrum taken as a template for the night. Subsequently, we refined the wavelength solution by utilizing an Earth's atmospheric transmission spectrum generated using the Sky Model Calculator.\footnote{\url{https://www.eso.org/observing/etc/bin/gen/form?INS.MODE=swspectr+INS.NAME=SKYCALC}}
For the remaining analysis, our focus was solely on order \#39, which includes the \Hei triplet. The magnitude of the wavelength calibration refinements is $\sim$0.56 \kms, approximately one fifth of a pixel.

To account for the Earth's atmosphere contamination, we utilised the ESO software {\tt\string Molecfit} \footnote{\url{http://www.eso.org/sci/software/pipelines/skytools/molecfit}}  \citep{2015A&A...576A..77S, 2015A&A...576A..78K} to correct for the transmission telluric lines, with a particular focus on the H$_2$O lines at approximately 1083.51 nm. The GIANO-B nodding acquisition mode automatically corrects for OH emission lines. However, in our previous works \citep[e.g.,][]{Guilluy2023, Guilluy2024}, we observed that the A-B subtraction could leave residuals at the position of the deepest OH doublet (around 1083.43 nm) due to variations in atmospheric seeing during observations. To address this, we masked any potential residuals in the final transmission spectra, as is described in \citet{Guilluy2023}.

\begin{figure*}
    \includegraphics[width=\linewidth]{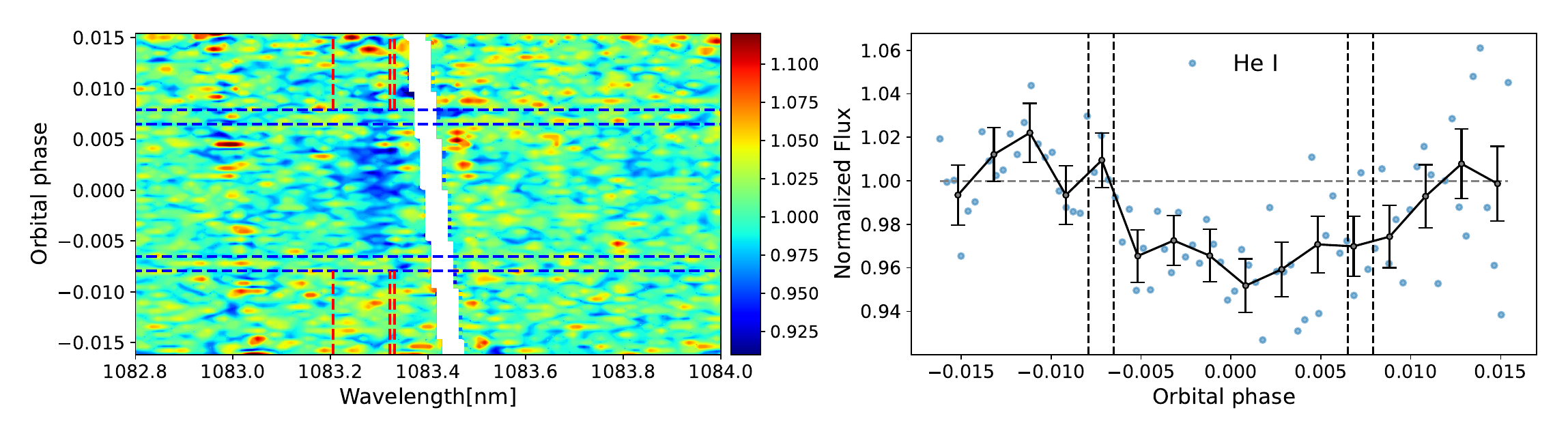}
    \caption{Analysis results for the \Hei triplet.
    Left panel: 2D maps of transmission spectra in the planet rest frame in the region of the \Hei triplet, as a function of wavelength and planetary orbital phase. The contact points, t$_1$, t$_2$, t$_3$, and t$_4$, are marked with horizontal blue lines. The regions affected by OH contamination are masked.  Some residuals are left at the position of the Si line ($\sim$1083 nm). This is due to the depth of the line, which can cause difficulties in spectral extraction \citep[see, e.g., ][]{Krishnamurthy2023}. Dashed red lines mark the position of the \Hei lines. Right panel: Light curve in the planet rest frame computed in a bandpass of 0.075~nm (equivalent to $\sim$ 20 \kms). Black points were computed with a phase bin of 0.002. Vertical dashed black lines indicate the position of the transit contact points, t$_1$, t$_2$, t$_3$, and t$_4$.  }
\label{HE}
\end{figure*}

\begin{figure*}
    \centering
    \includegraphics[width=0.86\linewidth]{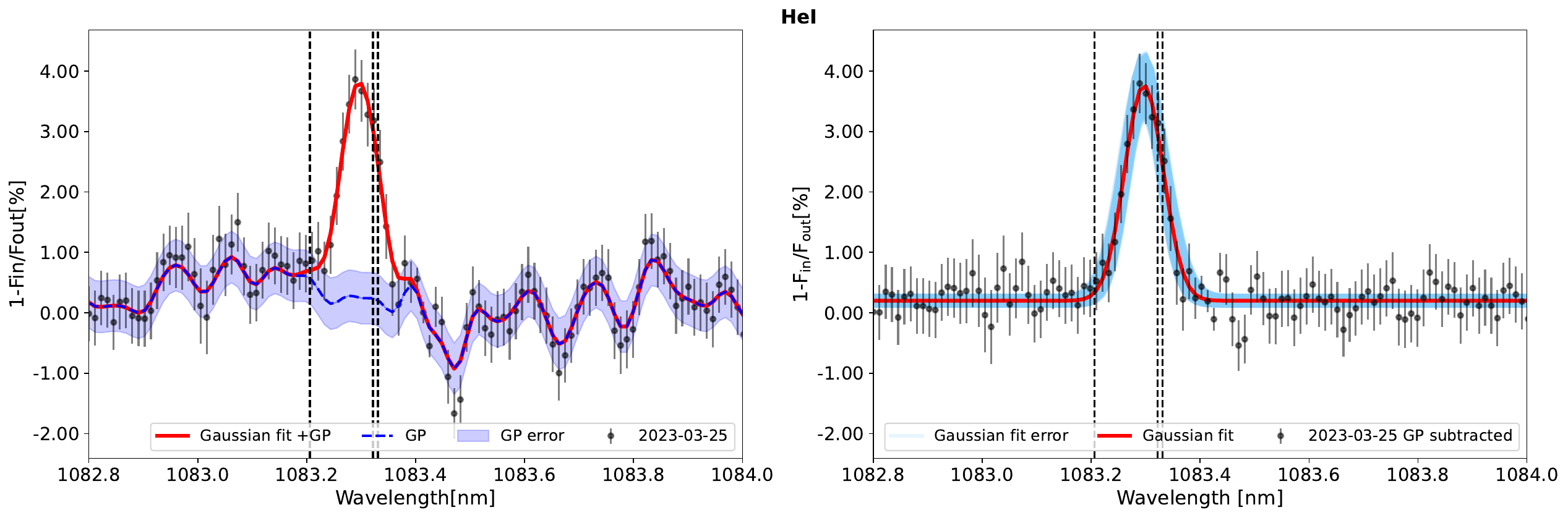}
    \includegraphics[width=0.85\linewidth]{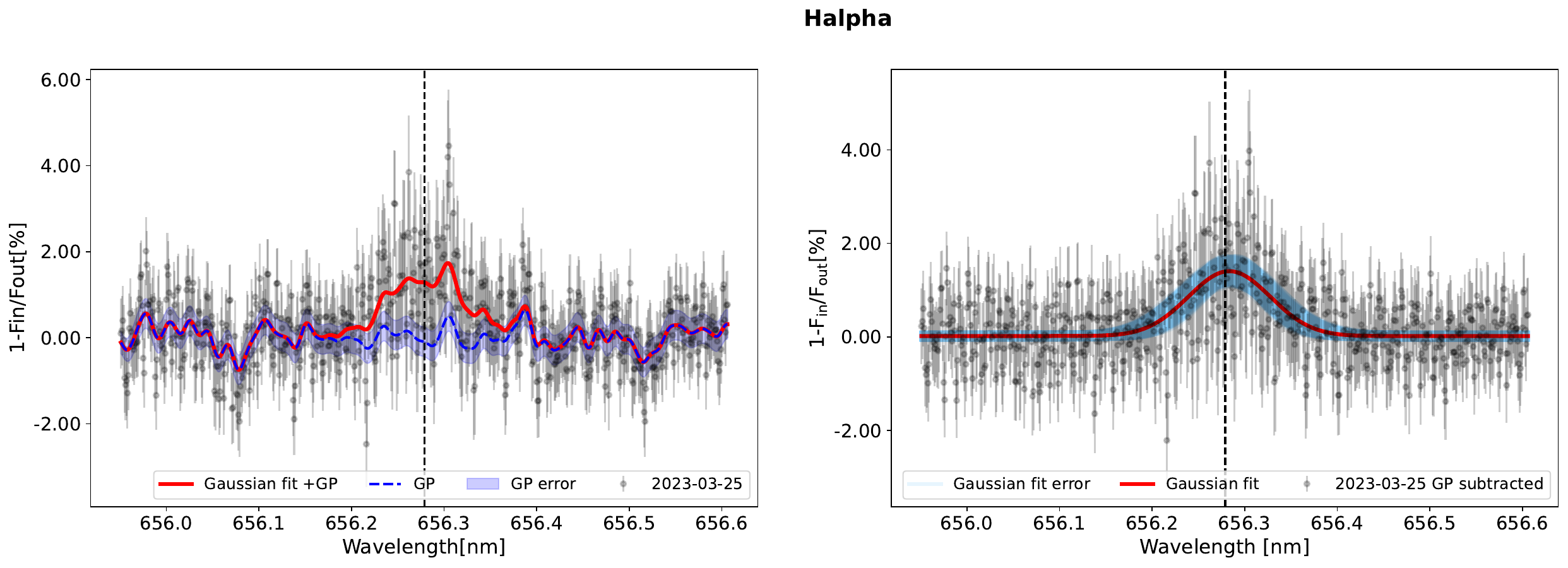}
    \includegraphics[width=0.85\linewidth]{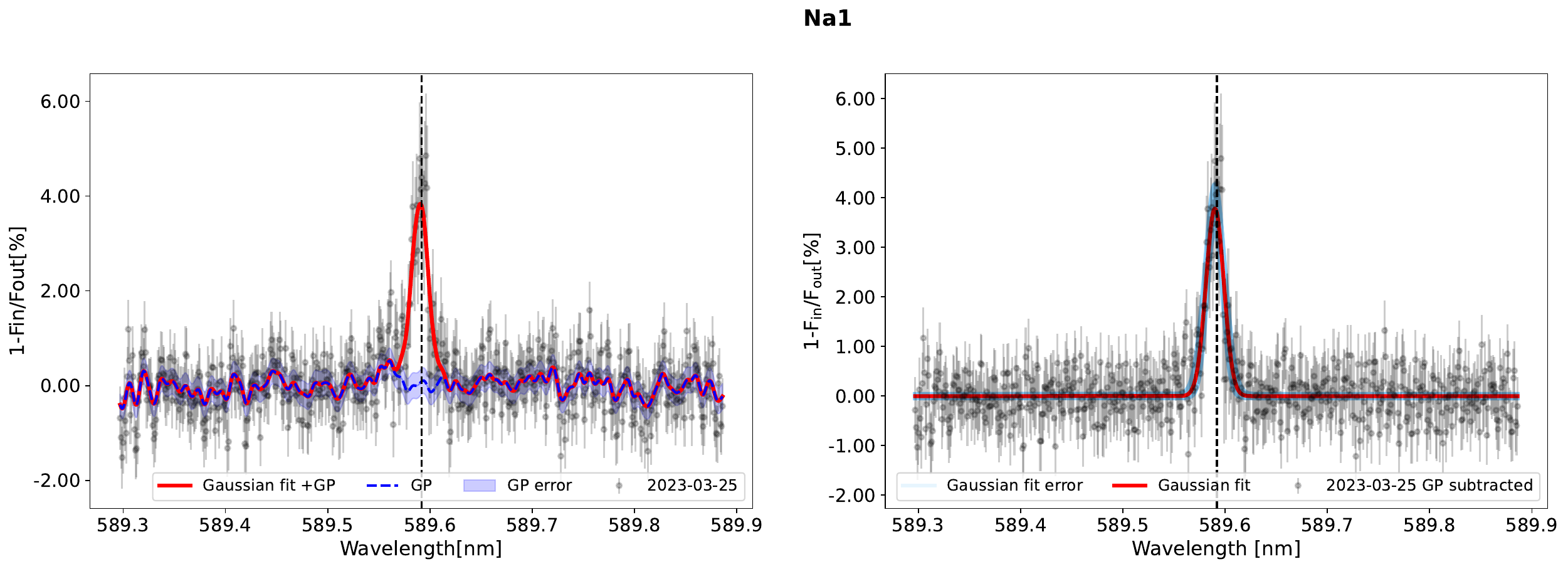}
    \includegraphics[width=0.85\linewidth]{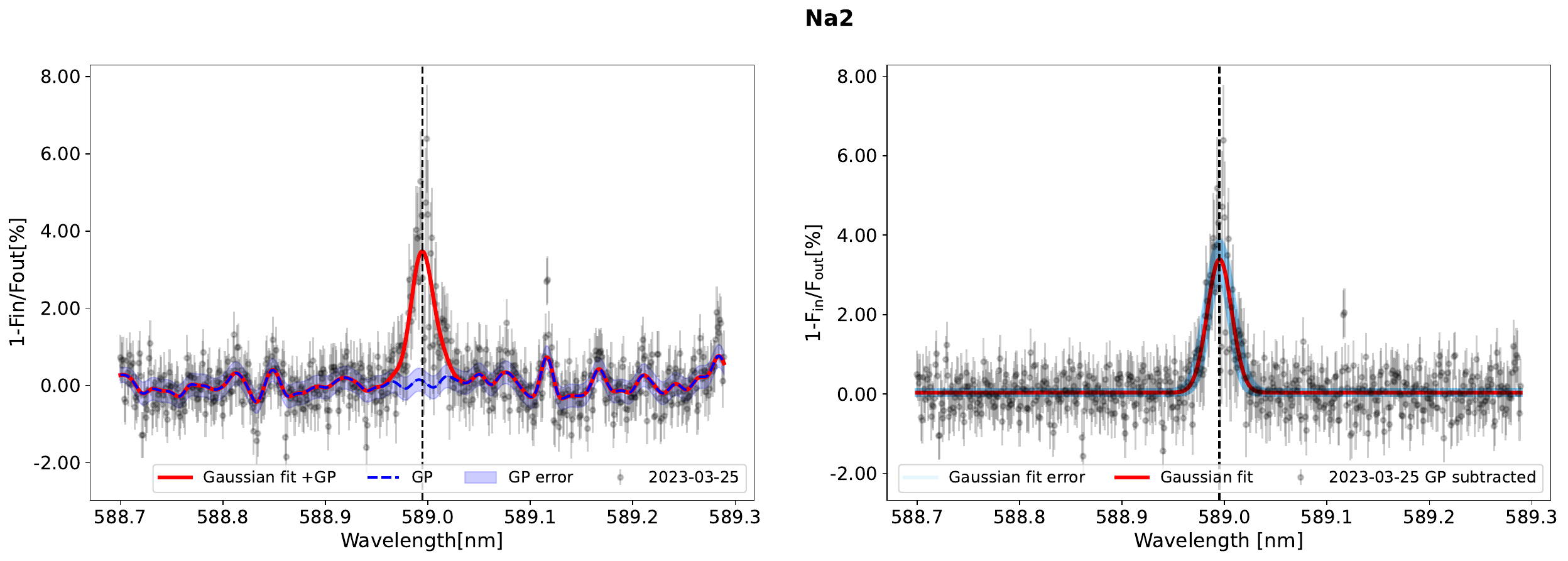}
       \caption{Transmission spectra of He I triplet and H$\alpha$ and Na I doublet after GP correction.  For each investigated line, the GP correction is shown. Left panel: Transmission spectrum centred on the line (in the
planet rest frame) with the GP regression model overplotted, along with the 1$\sigma$ uncertainty intervals (in blue), and the Gaussian+GP model (in
red). Right panel: Final transmission spectrum after removing the GP model. Vertical dotted black lines indicate the position of the investigated line.
 The error intervals for the Gaussian fit were computed by displaying 1000 Gaussian fits within the 1$\sigma$ uncertainties of the derived parameters, spanning the 16\%-84\% quantiles.}
 \label{GP+MCMC}
\end{figure*}

To separate the stellar contribution from the potential planetary signal, we performed transmission spectroscopy on the GIANO-B spectra according to the following approach. To start with, we moved the spectra to the star's rest frame and then we normalised the spectra to unity by median division, excluding the spectral region around the \ion{He}{I} triplet. Following that, we generated a master stellar spectrum, $S_\mathrm{out}(\lambda)$, by averaging the out-of-transit spectra (i.e. with an orbital phase smaller than t$_1$ or greater than t$_4$, the two ingress and egress contact points). We performed a visual comparison between $S_\mathrm{out}(\lambda)$ and a master-in spectrum, which was derived by averaging the in-transit between t$_2$ and t$_3$ (that is, the contact points of total transit). An absorption feature is readily discernible in the stellar spectrum, precisely coinciding with the position of the \ion{He}{I} triplet (Fig.~\ref{real_spectra}).
To better investigate this feature, we derived individual transmission spectra, $T\mathrm{(\lambda,i)}$, by dividing each spectrum by $S_\mathrm{out}(\lambda)$. Finally, we linearly interpolated the transmission spectra in the planet's rest frame. The left panel of Fig.~\ref{HE} displays the 2D transmission spectroscopy map, while the right panel shows the \Hei spectroscopic light curve computed within a passband of 0.075 nm centred on the peak of excess
absorption in the planet rest frame \citep[][]{Allart2019}. 
To quantify the contrast, $c$, of the extra absorption observed at the position of the \Hei triplet, we employed the differential evolution (DE) Markov chain Monte Carlo (MCMC) method \citep[][]{TerBraak2006}. This involved fitting a Gaussian profile to the mean in-transit transmission spectrum, computed by averaging the 2D maps between the transit contact points, t$_2$ and t$_3$. We fitted the peak position, the full width at half maximum (FWHM), the contrast value ($c$), and an offset for the continuum. Correlated noise in the transmission spectrum was accounted for using Gaussian process (GP) regression within the same DE-MCMC framework, employing a covariance matrix described by a squared exponential kernel \citep{Guilluy2024, 2024arXiv240403317S}. Additionally, uncorrelated noise was taken into account through a jitter term, $\sigma_\mathrm{j}$.
The correction with the GP regression model is illustrated in Fig.~\ref{GP+MCMC}, with detailed posterior distributions provided in Fig.~\ref{Cornerplots}.
The best-fit parameters obtained from the DE-MCMC Gaussian analysis are summarised in Table~\ref{tab_result_combined}. Parameter values and their 1$\sigma$ uncertainties were determined from the medians and the 16\%-84\% quantiles of their posterior distributions.

Table~\ref{tab_result_combined} also reports the effective \Hei radius \citep[e.g.,][]{Chen2018} that would produce the observed absorption contrast, $c$. We then normalised it to the atmospheric $H_\mathrm{eq}$ to compute the quantity, $\delta_\mathrm{R_P}/H_\mathrm{eq}$ \citep{Nortmann2018}, which represents the number of scale heights probed by the atmosphere in the spectral range under consideration. Here, $H_\mathrm{eq}$ was computed by assuming a mean molecular weight of 1.3. \citep{Fossati2022}.



\begin{table*}[h]
\caption{Best-fit parameters.}
\small
        \centering
        \begin{tabular}{c | c | c | c | c | c | c | c }
                \hline \hline
                Line & \multicolumn{2}{|c|}{Peak position} & Contrast $c$ &  R$_\mathrm{eff}$ & FWHM & Significance &  $\delta_\mathrm{R_P}$/H$_{\mathrm{eq}}$  \\
     &           [nm] & [\kms]     &         [\%]   & [Rp]      & [nm]   & [$\sigma$] &          \\ 
    \hline
\Hei & 1083.2969 $^{+ 0.0052 }_{ -0.0042 }$ & -8.09 $^{+ 1.44 }_{ -1.17}$ & 3.57 $^{+ 0.51 }_{ -0.50 }$ & 2.33 $\pm$ 0.02 &  0.084 $^{+ 0.016 }_{ -0.012 }$ & 6.9 &  79$\pm$9                   \\
H$\alpha$ & 656.2834 $^{+ 0.0094 }_{ -0.0088 }$& 1.94 $^{+ 4.42 }_{ -4.18}$ & 1.39 $^{+ 0.28 }_{ -0.28 }$ & 1.65 $\pm$ 0.11 & 0.104 $^{+ 0.028 }_{ -0.022 }$ & 5.0   & 39$\pm$7 \\
NaD1 & 589.5900 $^{+ 0.0011 }_{ -0.0011 }$ & -0.98 $^{+ 0.54 }_{ -0.55}$ & 3.79 $^{+ 0.50 }_{ -0.47 }$ & 2.38 $\pm$ 0.13 & 0.021 $^{+ 0.004 }_{ -0.003 }$ & 7.9  & 82$\pm$12 \\
NaD2 & 588.9952 $^{+ 0.0017 }_{ -0.0017 }$ & 0.14 $^{+ 0.85 }_{ -0.82}$ & 3.36 $^{+ 0.46 }_{ -0.44 }$ & 2.28 $\pm$ 0.12 & 0.028 $^{+ 0.004 }_{ -0.004 }$ & 7.4  & 75$\pm$11 \\                
\hline
        \end{tabular}
        \tablefoot{From left to right: the investigated spectral line, the peak position (both in air wavelength and velocity respect to the planetary reference frame), the excess of absorption, $c$, the corresponding effective radius, the FWHM, the significance of the detection, and the ratio between the equivalent height of the atmosphere and the atmospheric scale height.}
        \label{tab_result_combined}
\end{table*}

\subsubsection{HARPS-N visible region}
\label{vis_method}
We obtained the transmission spectra of the individual lines using a method similar to the one outlined in \cite{wyttenbach2015spectrally}, and thus compared spectra during and outside of transit. The raw data had already been processed using version 3.7 of the HARPS-N Data Reduction Software \citep[DRS;][]{2002A&A...388..632P}. At this point, the 1D spectra we employed included the stellar signal, the planetary signal, and telluric contamination, all presented in the solar system barycentric reference frame with the wavelength information provided in the air reference frame.

Correcting for telluric absorption is crucial, especially in the red region of the HARPS-N spectra, where the Earth's atmosphere shows significant H$_2$O and O$_2$ absorption features. Figure~\ref{fig:telluric_correction} illustrates an example of removing telluric lines near the H$\alpha$ and Na {\sc i} doublet regions. The software {\tt\string Molecfit} provides telluric-corrected spectra in the solar barycentric reference frame. Consequently, we adjusted the spectra to the stellar reference frame by accounting for the star's systemic radial velocity, v$_\mathrm{sys}$. The telluric-corrected spectra were then normalised to a common continuum level within a narrow wavelength range around each absorption line of interest. Before transitioning to the planetary reference frame, we must consider the combined effects of the RME \citep{rossiter1924detection, mclaughlin1924some} and CLV, which result from stellar rotation and limb darkening, respectively.
We addressed the RME+CLV effects by employing a similar approach to the one outlined by \cite{yan2017effect}, utilising stellar models. Specifically, we used ATLAS9 stellar models \citep{kurucz1992model, 2005MSAIS...8...73K, 2014dapb.book...39K, 2017ascl.soft10017K} to generate the disc-integrated stellar model based on the system parameters listed in Table \ref{tab_par}. We computed spectra for both a non-rotating star and a rotating one. To determine the portion of the stellar disc obscured by the planet at each phase (referred to as the obscured region), we used {\tt\string PyLightCurve} \citep{tsiaras2016new} to calculate the planet's transit path. We then created a grid with a pixel size of 0.01 R$_*$ $\times$ 0.01 R$_*$ to approximate the stellar disc and calculated the stellar flux for the obscured region at each phase. The intensity of each pixel was determined based on the limb darkening angle and the radial velocity shift relative to the stellar rotation axis. Using the ATLAS9 models, we computed the spectrum of the obscured region by summing the spectra of each pixel, each appropriately shifted by its own radial velocity. We derived the limb darkening coefficients for our stellar model using the {\tt\string ExoTethys} \citep{morello2020exotethys} function {\tt\string Sail} with a quadratic limb darkening law, consistent with ATLAS9. The integrated flux, $F_{\lambda}$, corresponds to the spectrum of the non-rotating model. Rotational broadening was obtained by summing the contributions of each pixel, each shifted by its own radial velocity. The observed master-out spectrum is broadened by stellar rotation, so we used the rotationally broadened model to normalise the continuum of the non-rotating model. Finally, we subtracted the modelled obscured region from the master-out spectra and divided all spectra by this result. 
Once the spectra in the stellar reference frame had been corrected, we shifted all the in-transit spectra to the planet reference frame by correcting for its radial velocity profile in a circular orbit scenario using the parameters listed in Table \ref{tab_par}. The final transmission spectrum of each line was calculated as the error-weighted average of all the full in-transit spectra (t$_2$ - t$_3$) in the planet reference frame.
We then performed a Gaussian fit, returning the absorption depth, the FWHM, and the velocity shift with respect to the line wavelength corresponding to a zero radial velocity. In the beginning, we used the planetary radius to represent the size of the obscured region of the stellar disc. We then translated the line absorption depths, returned by the Gaussian fit, into planetary radii and  we used these values to repeat the RME + CLV removal with an increased size that accounts for both the planetary radius and the atmospheric extension. We repeated the loop, retrieving a new depth expressed in planetary radii, until the threshold convergence (0.001 Rp) was reached. 
Once we found the optimal value for the planet+atmosphere region of the stellar disc obscured, we used the same DE-MCMC+GP tool employed in the GIANO-B analysis to fit our final corrected transmission spectra. Figure~\ref{GP+MCMC} shows the correction with the GP regression model, while the posterior distributions are provided in Fig.~\ref{Cornerplots}.
The best-fit parameters obtained from the DE-MCMC Gaussian analysis are reported in Table~\ref{tab_result_combined}. As for the \Hei contrast, we derived the corresponding effective radius for both H$\alpha$, and the \ion{Na}{I} doublet.
As the correlated noise is weaker in HARPS-N spectra compared to the GIANO-B ones, we assessed the necessity of the GP regression model correction using the Bayesian information criterion \citep[BIC,][]{kass_bic}. We considered the Gaussian+GP model to be the favourite compared to the simple Gaussian model (which also considered uncorrelated jitter) only if accompanied by a lower BIC and a difference in BIC ($\Delta$BIC) > 10, along with a Bayesian evidence >150. For all the three investigated lines (i.e. H$\alpha$, NaD1, and NaD2), the Gaussian+GP model emerged as the preferred choice. We provided the $\Delta$BIC and the corresponding Bayesian evidence in Table~\ref{BIC_table}. We computed the spectroscopic light curves similarly to the \Hei ones, considering a width of 20 \kms.

\subsection{Cross-correlation with templates}
To optimise the detection of certain elements with hundreds of weak lines, the cross-correlation function (CCF) is more suitable than transmission spectroscopy analysis \citep{brogi2012signature,hoeijmakers2019spectral,borsa_rot}, especially in the case of individual nights, since it allows one to boost the S/N. We used the CCF method in the visible band, searching for any exoplanetary signal. Before applying the CCF technique to our HARPS-N data, we employed the same telluric profile retrieved from the HARPS-N s1d spectra using {\fontfamily{pcr}\selectfont Molecfit}, and we used it to correct the HARPS-N e2ds orders from telluric lines for each exposure \citep{Hoeijmakers_2020}. To improve the accuracy of the CCF analysis, we chose to omit the initial five orders and the last one from all exposures due to their low S/N. We constructed the mono-dimensional spectra from normalised telluric-corrected e2ds orders. These 1D spectra were then cross-correlated with the templates at 2000K, 2500K, and 3000K provided by \cite{2023A&A...669A.113K} after convolving them with the HARPS-N resolution (R $\sim$ 115,000). Since these templates were derived using the Sun's radius as a reference, before any CCF computation, we multiplied them by the ratio $(R_{\odot}/R_{*})^2$, as is stated in \cite{2023A&A...669A.113K}, using $R_{*}$ from Table \ref{tab_par}. The templates were shifted in radial velocities between [-200 \kms, +200 \kms] with a step of 1 \kms. The resulting CCF was then shifted in the stellar reference frame, interpolated in radial velocity between [-150 \kms, +150 \kms] maintaining the 1 \kms step, and divided by the mean of the out-of-transit (master-out). To achieve a more precise normalisation and address any imperfections, we divided the CCF by the median values across the exposures. To remove any residual fluctuations, we performed a Fourier transform, discarding all the frequencies beyond 100 \kms. 

In this way, we investigated the presence of all the species given by \cite{2023A&A...669A.113K}, including both neutral and ionised elements, with at least one absorption line in the visible band covered by our order selection. The resulting CCFs for some elements exhibit the characteristic Doppler shadow. We removed it following the \cite{Rainer_2021} approach: we shifted the data in the Doppler shadow reference frame and fitted each column in a region where the Doppler shadow fell with a fifth-degree polynomial. However, the specific geometry of the TOI-5398 system is such that the Doppler shadow and the planetary trace overlap \citep[see][for a similar scenario]{2022A&A...664A.121C,2024arXiv240403317S}. Therefore, in this scenario, the Doppler shadow signal may completely mask the potential planetary signal, and even if the Doppler shadow is modelled and removed, the potential planetary signal may also be eliminated. Figure \ref{fig:ccf_2000K} shows the CCF obtained with the Na {\sc i}, Ca {\sc i}, and Fe {\sc i} templates at 2000K. The Doppler shadow for these elements was modelled out for CaI and FeI.

\section{Results} \label{sec:results}
\subsection{Single line and cross-correlation analyses} \label{res:single}
By comparing the out-of-transit spectra with the in-transit ones, we were able to discern clear He {\sc i}, H$\alpha$, NaD1, and NaD2 absorption signals. 
Applying the framework described in Sect. \ref{sec:methods}, we estimated contrasts of the excess absorption of 3.57$^{+ 0.51 }_{ -0.50 }$\% (6.9$\sigma$), 1.39$^{+ 0.28 }_{ -0.28 }$\% (5.0$\sigma$), 3.79$^{+ 0.50 }_{ -0.47 }$\% (7.9$\sigma$), and 3.36$^{+ 0.46 }_{ -0.44 }$\% (7.4$\sigma$) for \Hei, H$\alpha$, NaD1, and NaD2, respectively. 
These values correspond to effective planetary radii of $\sim$2.33~Rp, $\sim$1.65~Rp, $\sim$3.81~Rp, and $\sim$2.28~Rp, the extensions of which are well below the planet's Roche lobe radius ($\sim$5.8Rp), computed using Eq. 2 from \citealt{1983ApJ...268..368E}. This indicates that the planetary atmosphere is not escaping due to the gravitational pull of the star, as is discussed in the next section.
Assuming a scale height ($H$) of $\sim$ 1106~km, and considering 1$\sigma$ uncertainties, we found that the \Hei atmosphere probes a number of scale heights varying between $\sim$ 70 and $\sim$ 88 $H$, while the H$\alpha$ atmosphere varies between 32 and 46 $H$, NaD1 between 70 and 96 $H$, and NaD2  between 64 and 86 $H$.
We found the \Hei absorption signal to show a clear blueshift, $\sim$~8\kms, while the optical species do not display either a significant blueshift or a redshift. \Hei and H$\alpha$ have similar FWHMs, larger than the Na {\sc i} doublet ones. For the \Hei and H$\alpha$ lines, our detections seem to be supported by the spectroscopic light curves, which hint at the presence of an \Hei tail that we discuss later. The Na doublet light curves do not show a clear transit signal, despite showing a decrease in flux in the second half of the transit. This is possibly related to the overlapping RME, as is discussed later.
We also investigated the presence of other metal species in the atmosphere, analysing a wide forest of lines belonging to Mg I, Ca I, Fe I, and Fe II, but unfortunately we did not detect any significant line. The cross-correlation of the extracted spectrum with \cite{2023A&A...669A.113K} templates did not return any detection of the investigated species. As was already stated, this framework is particularly challenging in our case due to the overlap of the possible signals with the Doppler shadow. This also affected the extraction of single lines in the visible spectrum of TOI-5398 b. An under-correction of RME may cause one to miss the signal, while an over-correction may lead to a false detection. We compared the results obtained applying the method described in Sect. \ref{vis_method} (already employed in \citealt{Guilluy2024} and \citealt{darpasub}) with the results obtained using SLOPpy \citep{2022A&A...667A..19S}, which employs a different method to model the RME+CLV, as is described in \ref{app:sloppy}.

\subsection{Comparison with ATES predicted spectra} \label{sec:ates}
To support the interpretation of our \Hei observations, we conducted simulations of the atmospheres of the two planets and their respective \Hei absorptions using the 1D hydro-photoionisation self-consistent code ATES \citep{Caldiroli_2021} and the new transmission probability module \citep[TPM,][]{Biassoni_2023}. The ATES code derives values pertaining to the 1D atmospheric radial profile such as temperature, pressure, density, and velocity, along with the mass loss rate, $\dot{M}$. 

For these simulations, we used the planetary parameters listed in Table \ref{tab_par} for planets b and c, while the stellar spectrum  was calculated as follows. For the UV component, we adopted a stellar spectrum from the Phoenix library \citep{2013A&A...553A...6H}, to which we added the Lyman alpha emission following the prescriptions of \cite{2020ApJ...902....3L} and in the manner described in \cite{2022PSJ.....3....1L}. For the EUV and X-ray components, we produced a synthetic spectrum from a composite chromosphere-corona plasma emission measure distribution versus temperature, EMD(T), constructed as follows: the coronal part (log T > 5.5 K) was interpolated from the grid of EMDs provided by \cite{2018ApJ...862...66W} for a star with a surface X-ray flux of Fx = $1.5 \times 10^6$ \ergcms (corresponding to Lx = $1 \times 10^{29}$ erg~s$^{-1}$), while the chromospheric section (log T from 4.0 to 5.5 K) was approximated as a power law following \cite{2011A&A...532A...6S}, with the constraint that the Feuv/Fx ratio follows the scaling law indicated by \cite{2021A&A...649A..96J}. Then, the optically thin XUV spectrum (1.24 - 1700 A) was computed adopting the plasma emissivities from the CHIANTI v7.13 atomic database as in \cite{2023ApJ...951...18M}.

\begin{figure}
\centering
    \includegraphics[width=1.1\linewidth]{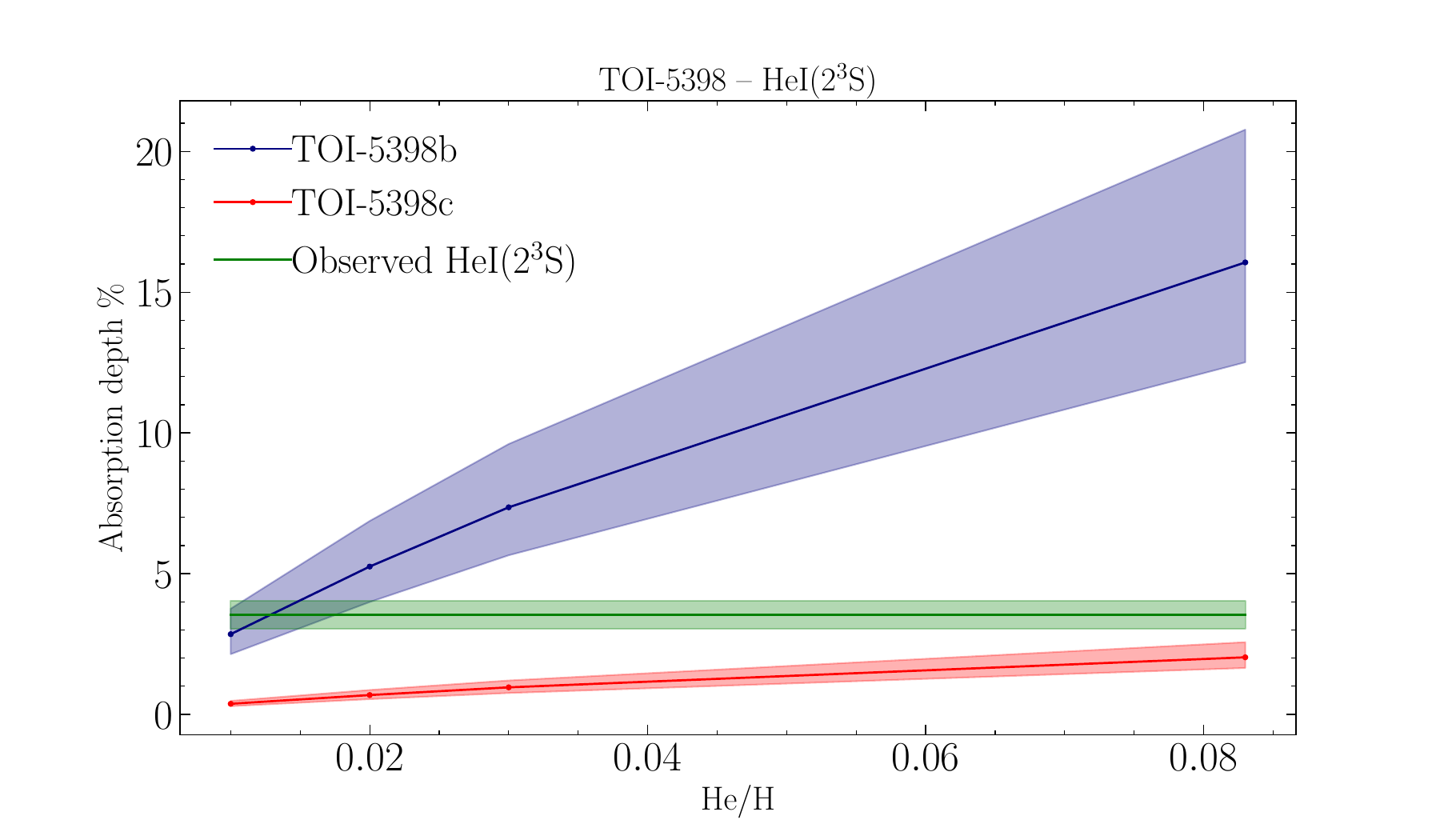}
    \caption{\Hei simulated absorption profiles for both planets changing the He/H number abundance. The green line corresponds to the \Hei observed signal with 1$\sigma$ error bars given in Table \ref{tab_result_combined}. The blue and red bands correspond to the simulations obtained for the two planets, varying the X-ray luminosity by a factor of two.}
    \label{fig:sim_He}
\end{figure}

We started our simulations for both planets assuming a constant He/H number abundance of $1/12$ alongside the overall atmosphere. This provided us $\dot{M}$ of ~$10^{11.70}$ \gs and $10^{11.44}$ \gs for planets b and c, respectively, indicating that nowadays the innermost planet, c, is affected less than planet b by the atmospheric escape effect. However, this does not exclude the possibility that in the past planet c may have experienced significantly greater atmospheric loss.
We investigated the \Hei absorption profiles for both planets using the TPM module. For this analysis, we firstly simulated the atmospheric profiles changing the He/H number abundance, using values of 0.01, 0.02, 0.04, and $1/12$. 
We then repeated the simulations, varying the XUV flux to include the enhanced stellar activity of a young star. TOI-5398 is a G-type star with an age comparable to the one of the Hyades cluster stars.
According to Fig. 4 of \cite{2005ApJS..160..390P}, in such stars, the typical X ray luminosity is centred at Lx = $1 \times 10^{29}$ erg~s$^{-1}$, which is the value we adopted, with a variation from the pre–main-sequence through the main sequence that is almost a factor of two. Therefore, we repeated the same simulations described above, varying the XUV luminosity by a factor of two. We used the results as a function of the He/H number abundance to estimate the uncertainties of the values retrieved with the nominal luminosity.
The resulting \Hei absorption profiles obtained with these different He/H number abundances for planet c are always below the observed signal, while for planet b, the simulations are consistent with the observed signal only with a He/H number abundance between 0.01 and 0.02 (see Fig. \ref{fig:sim_He}).

\section{Discussion}\label{sec:discussion}
Within the single-line analysis context, the \Hei triplet in the nIR, and the H$\alpha$ and Na {\sc i} doublet in the visible, are among the pivotal lines studied in exoplanetary atmospheres, thanks to their strength and to their role as markers of features such as photo-evaporation and stellar activity. 
The Na {\sc i} doublet has been detected in several hot exoplanets with temperatures ranging from 963 K in WASP-69 b \citep[e.g.,][]{2017A&A...608A.135C,2021A&A...656A.142K} to 3921 K in KELT-9 b \citep[e.g.,][]{2022MNRAS.514.5192L,darpasub}. Therefore, according to our detection, TOI-5398 b will be among the coldest (947 K) planets showing the Na {\sc i} doublet, to our knowledge. 
The \Hei and H$\alpha$ distribution has recently been discussed by \cite{2024arXiv240416732O}, who explored the high-resolution spectroscopy observations from CARMENES and GIARPS checking for \Hei and H$\alpha$ signals in 20 exoplanetary atmospheres. From their work, we can see that the joint detection of \Hei and H$\alpha$ has been observed only in two targets: WASP-52 b \citep{2020A&A...635A.171C,2022AJ....164...24K,2020AJ....159..278V} and HAT-P-32 b \citep{2022A&A...657A...6C, 2023SciA....9F8736Z}, with WASP-52 b being the only one with a detection of all three species detected in our study.

The height distribution of the species in our atmosphere suggests that \Hei and Na {\sc i} lie in an external atmospheric layer with respect to H$\alpha$.
H$\alpha$ forms through absorption from the n=2 level of neutral hydrogen, which is likely excited by the strong stellar Ly$\alpha$ line and/or by the photoionizing radiation of our young star followed by subsequent recombination and population of the n = 2 level as in solar-like stellar atmospheres \citep{1985ApJ...294..626C}. To justify the observed H$\alpha$ absorption, it is necessary for hydrogen to have a sufficient number of atoms in the excited level n=2. If we go too high in the planetary atmosphere, hydrogen finds itself in a less dense medium and has enough time to decay to the fundamental level n=1 before being able to absorb a photon in the H$\alpha$ line. In the case of the He I triplet, the situation is different because the lines form through absorption from a metastable level. Even if we are high in the atmosphere, the level decays over much longer timescales than those required for photon absorption in the lines. Moreover, in a less dense medium, the probability that the atom decays to the fundamental level due to collision is low, facilitating the process that produces the He I triplet.

Indeed, absorption of He I at 1083 nm is seen in certain cases, even at a great distance from the planet, along almost its entire orbit \citep[see][for example]{2023SciA....9F8736Z}. The sodium retrieved height is more challenging to explain, since the sodium atom has a low ionisation potential, and therefore, far from the planet, it risks being ionised. However, the absorption of the two doublet lines starts from the fundamental level, so the atom does not need a dense surrounding medium to bring it to an excited level through thermal collisions to produce the line. Therefore, if it does not ionise too much, some absorption in the Na {\sc i} doublet at a certain distance from the planet might be observable. In this case, we stress the mismatch between the results obtained with SLOPpy, where the Na {\sc i} doublet is found at 1.12 Rp. Since the species is not detected via cross-correlation with templates, the signal we detect may be due to a spurious signal arising from a residual of the RME+CLV contamination overlapping with the planetary signal. This hypothesis could explain the fact that the Na doublet lines do not show a clear transit signal in the light curve despite showing a flux decrease in the second half of the transit. Future observations and improvements in the RME+CLV modelling will be necessary to clarify this point.

The distribution of H$\alpha$ and Na {\sc i} resemble the same behaviour observed in WASP-52 b, where \cite{2020A&A...635A.171C} find both sodium lines lying at an effective radius larger than the H$\alpha$ one.
The extended atmosphere observed for our target may be explained taking into account the photo-evaporation induced by the host star.
In fact, planets close to their host star experience a strong irradiation that causes the exoplanetary atmosphere to heat up, and hence to expand. The larger thermal energy induces an increase in velocity, making the particles belonging to the atmospheric species reach the Roche lobe and escape from the planetary gravitational well. This is particularly relevant for younger stars, which generally have a larger XUV flux such as TOI-5398. \cite{mantovan_ross} evaluated photo-evaporation in the system during a period spanning from 3~Myr to 650~Myr. The two planets experienced a different evolutionary path, with planet b being less influenced by atmospheric escape, while planet c, on the other hand, experienced the loss of almost 60\% of its mass, resulting in a less extended atmosphere nowadays. At this point of its evolutionary path, planet c should show a less extended atmosphere, while planet b remains dominated by an extended envelope despite experiencing a currently larger mass loss.
The simulations we performed with ATES retrieve $\dot{M}$ of ~$10^{11.70}$ \gs and $10^{11.44}$ \gs for planets b and c, respectively, confirming that part of the outer layers of the atmosphere are currently undergoing atmospheric escape.
This scenario is ideal for transmission spectroscopy due to the larger extension of the scattering annulus surrounding the planet. 

From our results, we observe that TOI-5398 b shows an extended atmosphere that reaches $\sim$ 2.4 Rp, still well under the Roche lobe at $\sim$ 5.8 Rp, computed using Eq. 2 from \cite{1983ApJ...268..368E}. This does not question the photo-evaporation ongoing for the target, since we obtained the height from a geometrical point of view from the absorption depth of the lines detected. Since the latter is related to the point where the optical depth is one, the height does not represent an individual layer where the species lie, since it is distributed according to the density profile. Therefore, we are able to find particles overcoming the Roche lobe despite observing an absorption that peaks well below.

The past and current status of the two planets, along with the ATES models, should help us to address the possible contamination of planet c during the transit that we observed. In Fig. \ref{fig:obs}, we show that the two transits almost completely overlap and only three points of planet b are unaffected by planet c.
While it is impossible to completely exclude the uncertain contribution of planet c, its role in the observed signal seems to be secondary, as is jointly supported by the two diagnostics: planet c has a less extended atmosphere due to its past evaporation and the spectra retrieved by ATES (Fig. \ref{fig:sim_He}) confirm our suggestions, since the \Hei arising from the planet c is not able to justify the absorption we observe, even considering two times the nominal XUV flux. In particular, in order to match the \Hei triplet retrieved absorption, we should increase the XUV flux by more than ten times its value. On the other hand, planet b is expected to show hints of photo-evaporation and the ATES spectra manage to mimic the 3.57\% absorption with a He/H number fraction ranging between 0.01 and 0.02 . While retrieving the He/H number fraction is not the main aim of our comparison with the TSM module, we are able not only to evaluate the contribution of the two planets but also to discern a range of He/H number fraction as a function of the XUV flux adopted, which in a young star may vary. The use of models including the H$\alpha$ line may be able to break this degeneracy.

Observing the spectroscopic \Hei light curve in Fig.~\ref{HE}, we note that the signal seems to continue beyond T$_4$. This may be an indication of the presence of an extended atmosphere or some sort of cometary tail trailing the planet. The tail may be further proof of planet b being photo-evaporating. As further evidence, we have absorption of \Hei at ~8 \kms, which could originate from the cometary tail driven by stellar winds. However, the tail does not seem to be present in the other species' light curve, as it is in the \Hei one. This may arise from the lower number of points in the visible band due to a larger exposure time. 

\section{Conclusions} 
\label{sec:conclusions}
In this paper, we have performed a high-resolution nIR+optical transmission spectroscopy study of the Saturn-like planet TOI-5398\,b, analysing one transit collected on March 25 2023 with the GIARPS observing mode at the TNG. The same dataset has been used before by \cite{mantovan_ross} to constrain the orbital parameters and evaluate the evolutionary path of the system.

We detected the presence of metastable \Hei with the GIANO-B spectrograph and of a H$\alpha$ and Na {\sc i} doublet with HARPS-N employing single-line analysis. According to the ExoAtmospheres database,\footnote{\url{https://research.iac.es/proyecto/exoatmospheres/index.php}} this represents one of only a few simultaneous detections of He {\sc i} and an H$\alpha$ and Na {\sc i} doublet in the atmosphere of an exoplanet. 

The cross-correlation with templates did not return any detection, probably due to the overlapping RME, the low temperature of the planet, and the exiguous dataset.
In fact, the difficulty of extracting the planetary signal in the atmosphere of smaller planets is enhanced by the fact that we observed only one transit. As is discussed in \cite{Guilluy2024}, the study of individual lines across different nights is crucial to identify possible contamination due to stellar activity, and hence more transits will be necessary to further corroborate our detections. 

Despite us attributing the detected signal to planet b according to the evolutionary history of the system and to the ATES simulations, future individual observations of both planets will also be needed to definitively clarify the origin of the signals detected. 
Future observations with longer baselines would also help in investigating the possible presence of a cometary tail, along with 3D hydrodynamical simulations such as the ones performed in \cite{2015A&A...578A...6M} and \cite{2024A&A...683A.226C}.


\begin{acknowledgements}
We sincerely thank the Referee for their comments stressing some important factors and analyses that we did not consider initially, and that we think improved the quality of the manuscript.
The authors acknowledge financial contribution from PRIN INAF 2019 and from the European Union - Next Generation EU RRF M4C2 1.1 PRIN MUR 2022 project 2022CERJ49 (ESPLORA), PRIN MUR 2022 project PM4JLH (“Know your little neighbours: characterizing low-mass stars and planets in  the Solar neighbourhood”) and PRIN MUR 2022 (project No. 2022J7ZFRA,  EXO-CASH).
The authors acknowledge the support of the ASI-INAF agreement 2021-5-HH.0 and the project “INAF-Astrofisica Fondamentale GAPS2" and INAF GO Large Grant 2023 GAPS-2.
GMa acknowledges support from CHEOPS ASI-INAF agreement n. 2019-29-HH.0. Part of the research activities described in this paper were carried out with contribution of the Next Generation EU funds within the National Recovery and Resilience Plan (PNRR), Mission 4 - Education and Research, Component 2 - From Research to Business (M4C2), Investment Line 3.1 - Strengthening and creation of Research Infrastructures, Project IR0000034 – “STILES - Strengthening the Italian Leadership in ELT and SKA”.
The authors acknowledge the Italian center for Astronomical Archives (IA2, https://www.ia2.inaf.it), part of the Italian National Institute for Astrophysics (INAF), for providing technical assistance, services and supporting activities of the GAPS collaboration.\\
\end{acknowledgements}

\bibliographystyle{aa} 
\bibliography{aa51237-24corr}

\begin{appendix}
\onecolumn
\section{Additional figures and tables}
\begin{figure}[h!]
    \includegraphics[width=\linewidth]{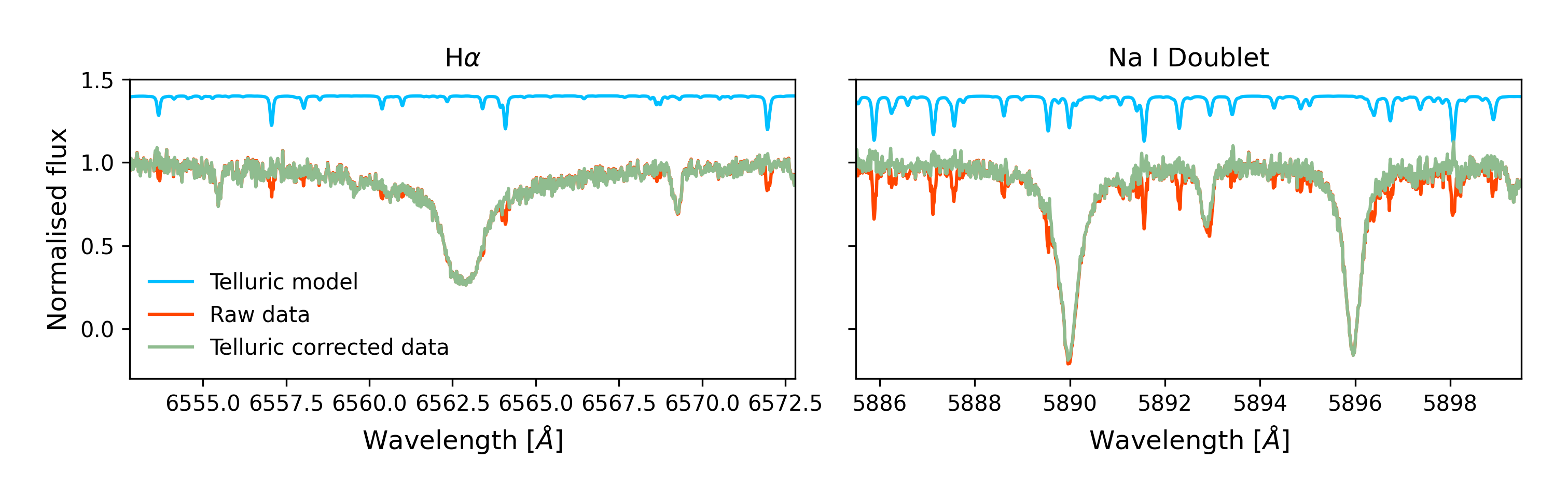}
    \caption{Example of telluric correction for the H$\alpha$ and Na {\sc i} doublet regions. After normalisation, the DRS processed data (in orange) was telluric corrected with {\tt\string Molecfit} using the telluric model in blue, here shifted upwards for clarity.}
    \label{fig:telluric_correction}
\end{figure}
\FloatBarrier

\begin{figure*}[h]
    \includegraphics[width=\linewidth]{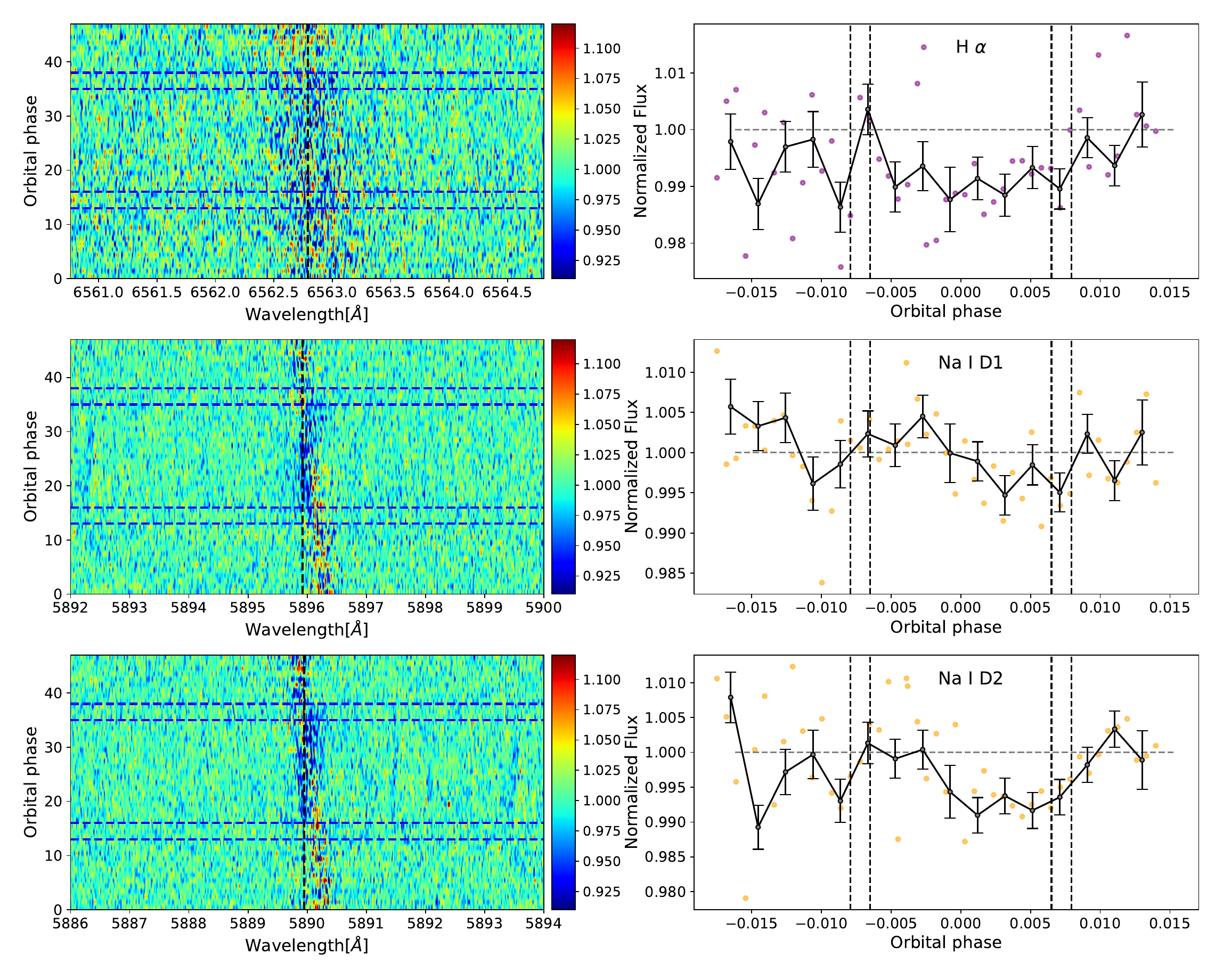}
    \caption{Same as Fig. \ref{HE} but for the lines detected in the visible range. The dashed black lines represent the wavelength position of the line.}
\label{vis}
\end{figure*}

\begin{figure*}[h]
    \includegraphics[width=0.47\textwidth]{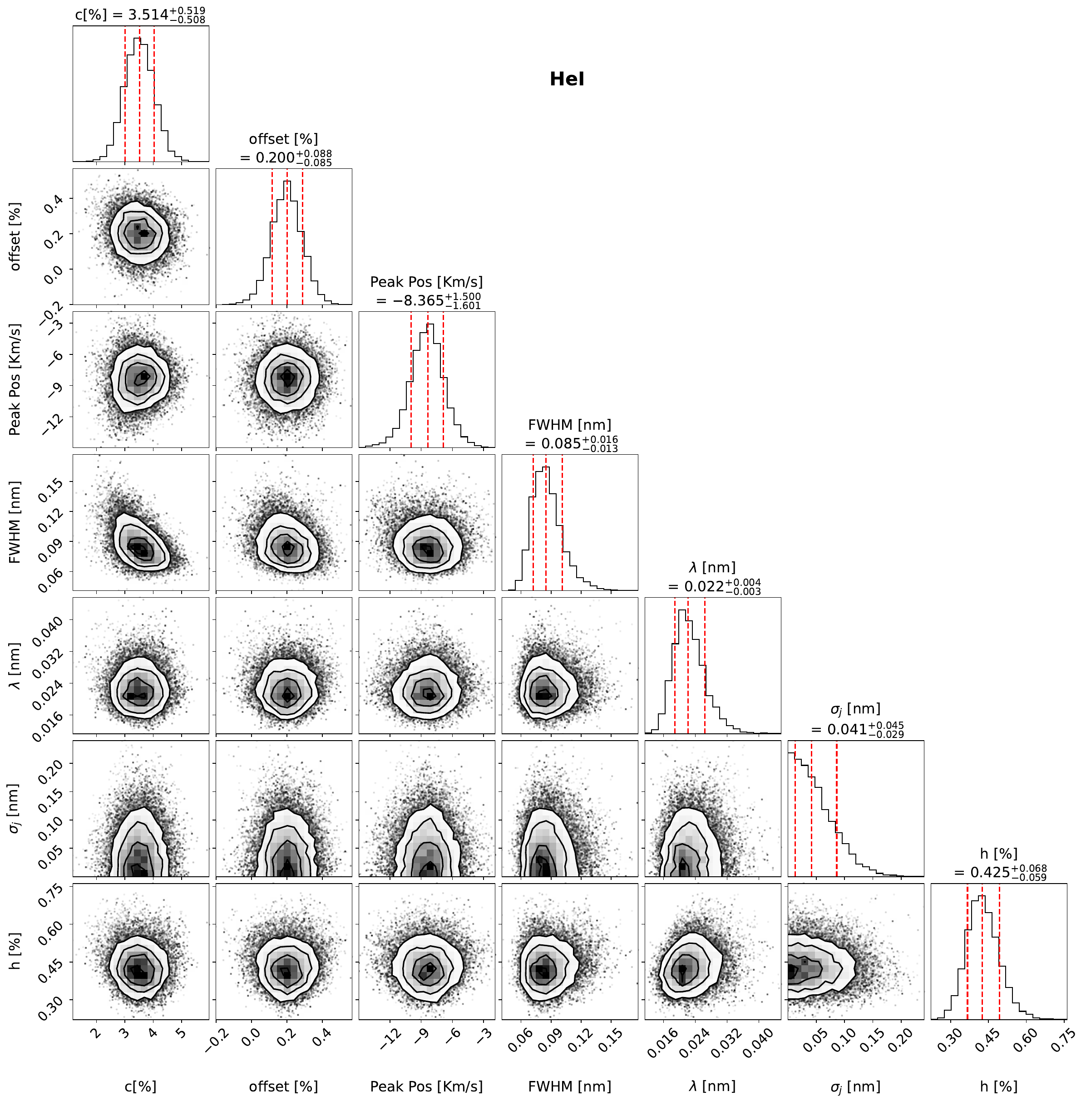}
    \includegraphics[width=0.47\textwidth]{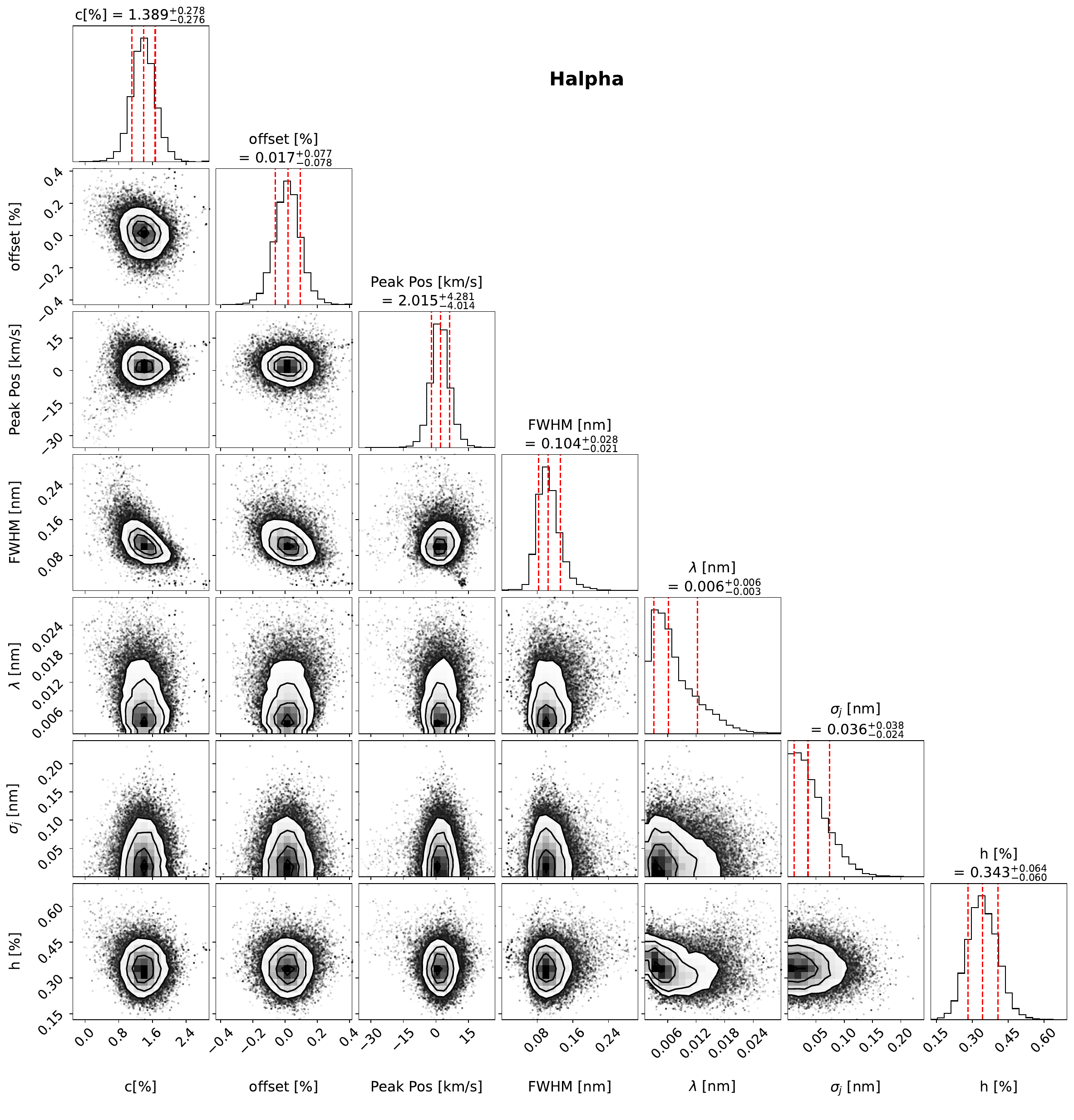}\\
    \includegraphics[width=0.47\textwidth]{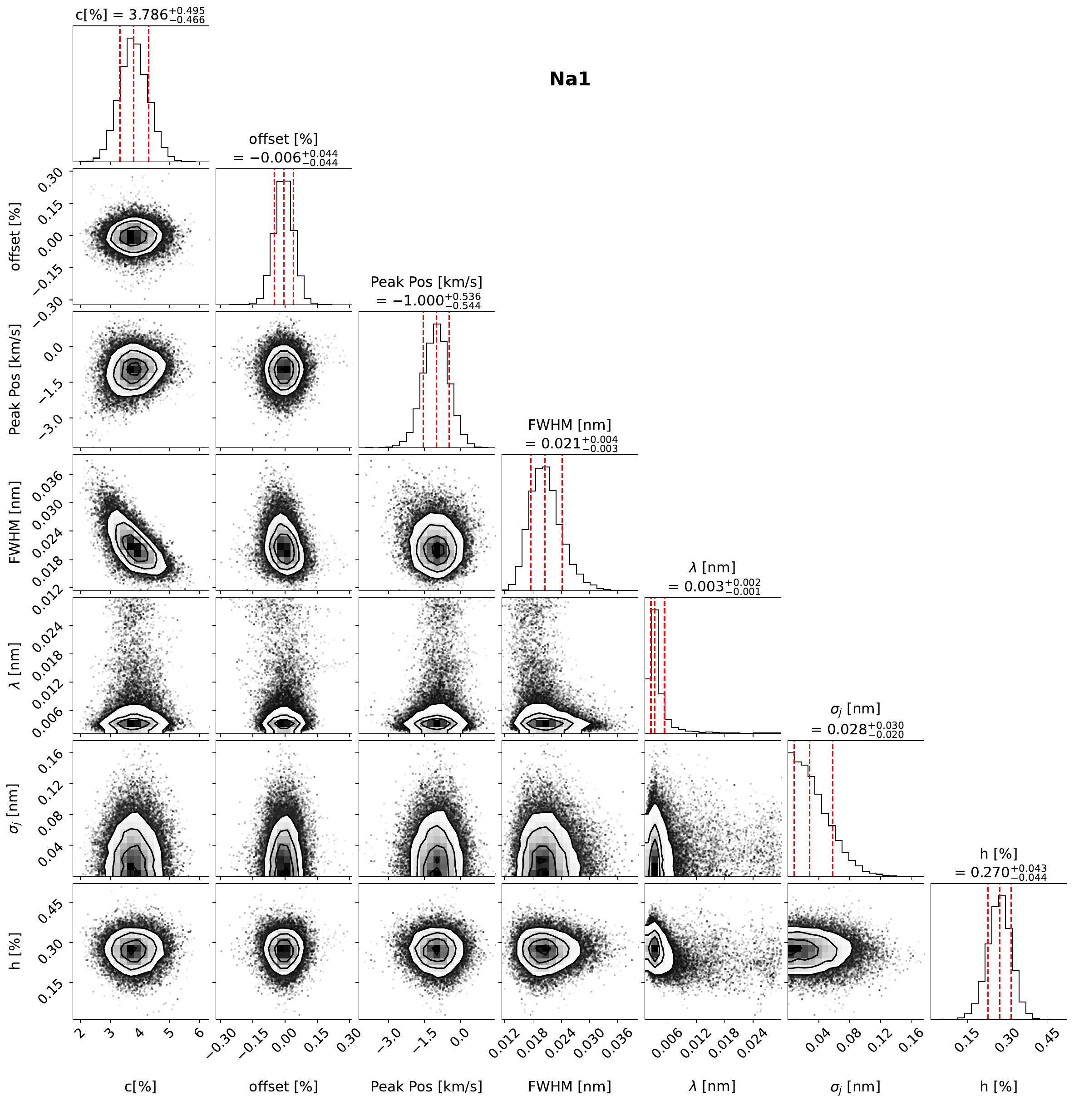}
    \includegraphics[width=0.47\textwidth]{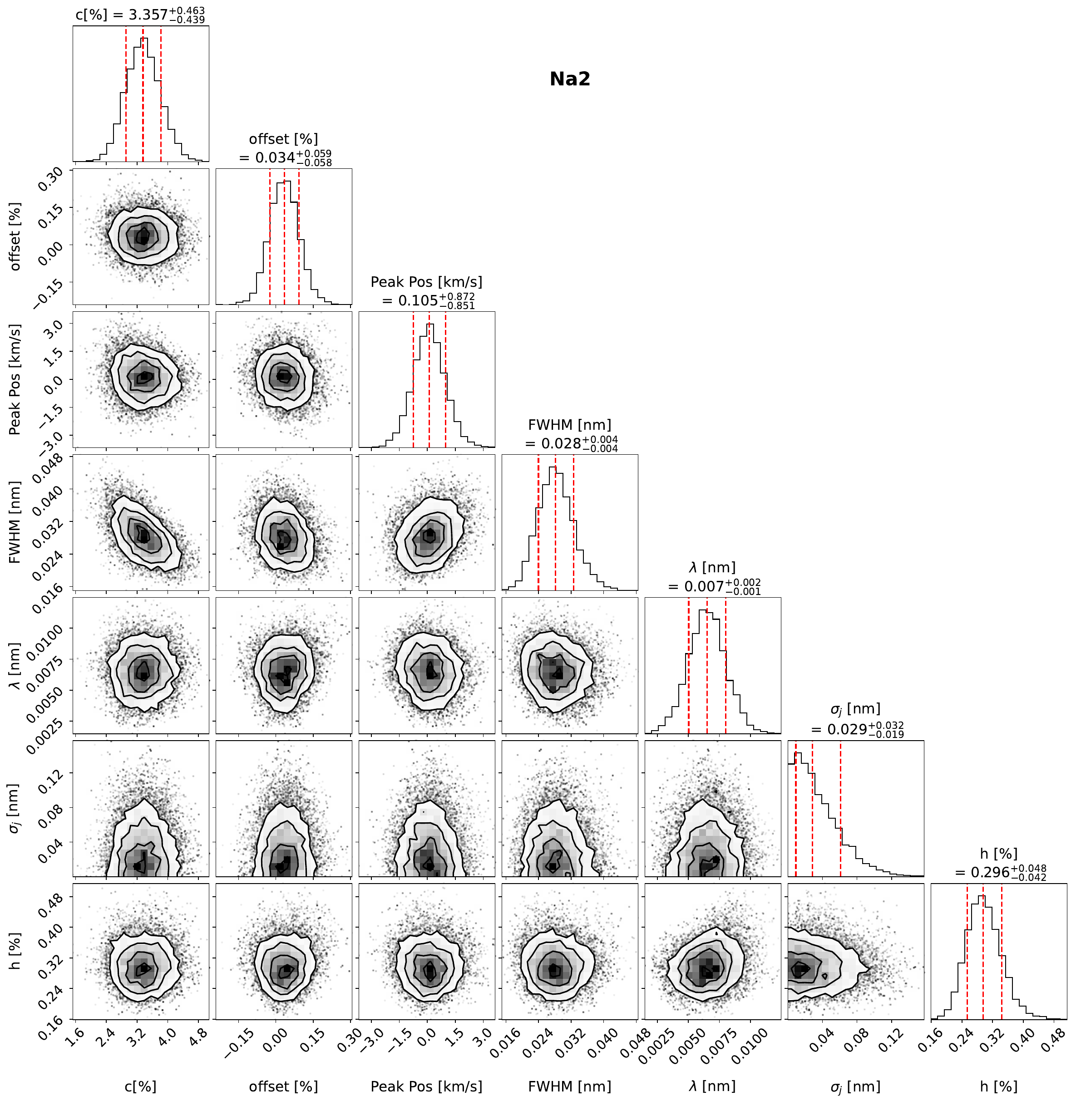}
    \caption{Posterior distribution of the investigated parameters in the DE-MCMC analysis for the four investigated lines. The excess of absorption $c$ [\%], offset [\%], peak position, and FWHM correspond to the parameters we used in the Gaussian fit, while the jitter term $\sigma_\mathrm{j}$, the semi-amplitude of the correlated noise $h$, the correlation length $\lambda$ were used to parametrise the SE kernel within the GP. }
    \label{Cornerplots}
\vspace{1cm}
\centering
  \captionof{table}{Gaussian+GP and Gaussian model comparison via BIC in HARPS-N spectra.}
  \begin{tabular}{c|c|c}
  \hline\hline
  Line & $\Delta$BIC & Bayesian evidence \\ 
  \hline
  H$\alpha$    & 13.3 & 783.8 \\
  NaD1     & 14.7 & 1518.6 \\
  NaD2     & 26.8 & 654277.3\\  
  \end{tabular}
  \label{BIC_table}
\end{figure*}

\begin{figure*}[h]
    \includegraphics[width=\linewidth]{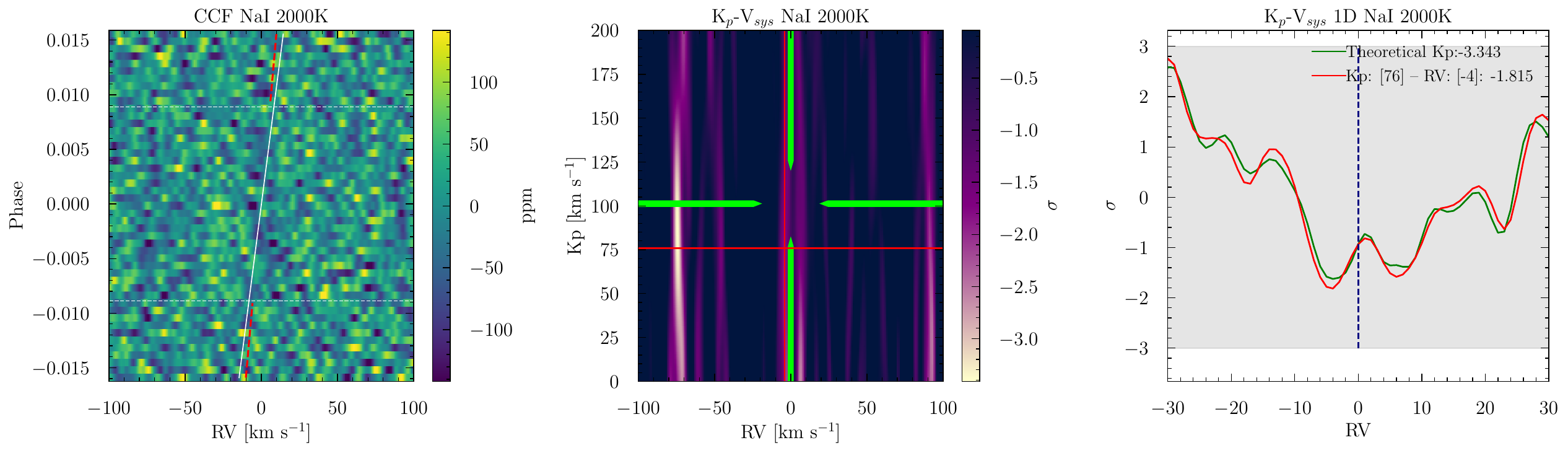}
    \includegraphics[width=\linewidth]{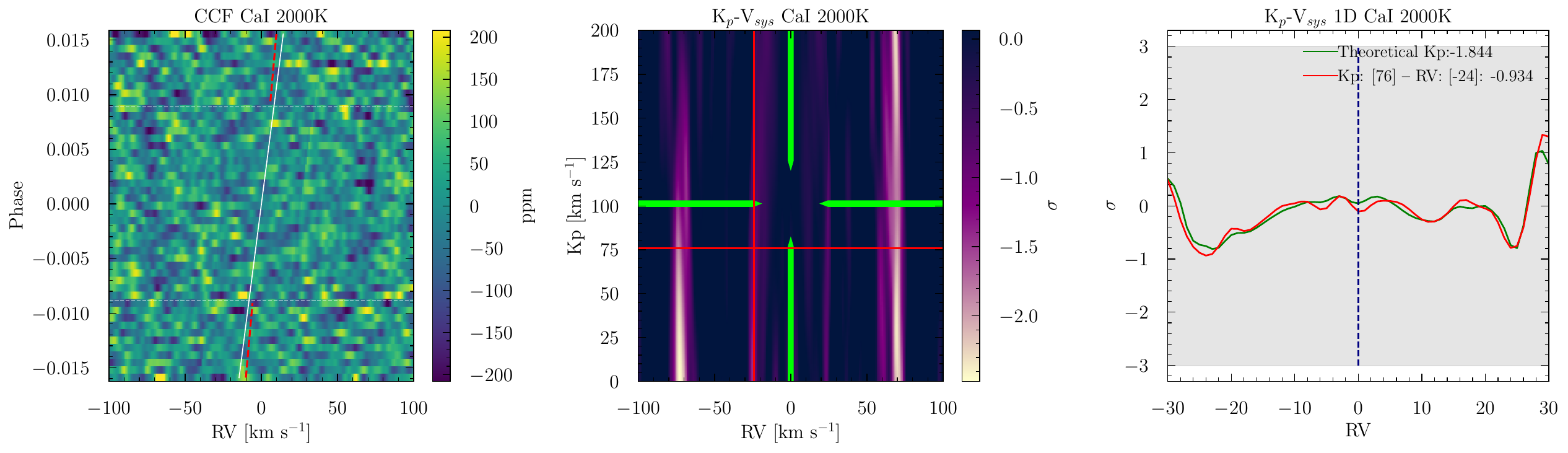}
    \includegraphics[width=\linewidth]{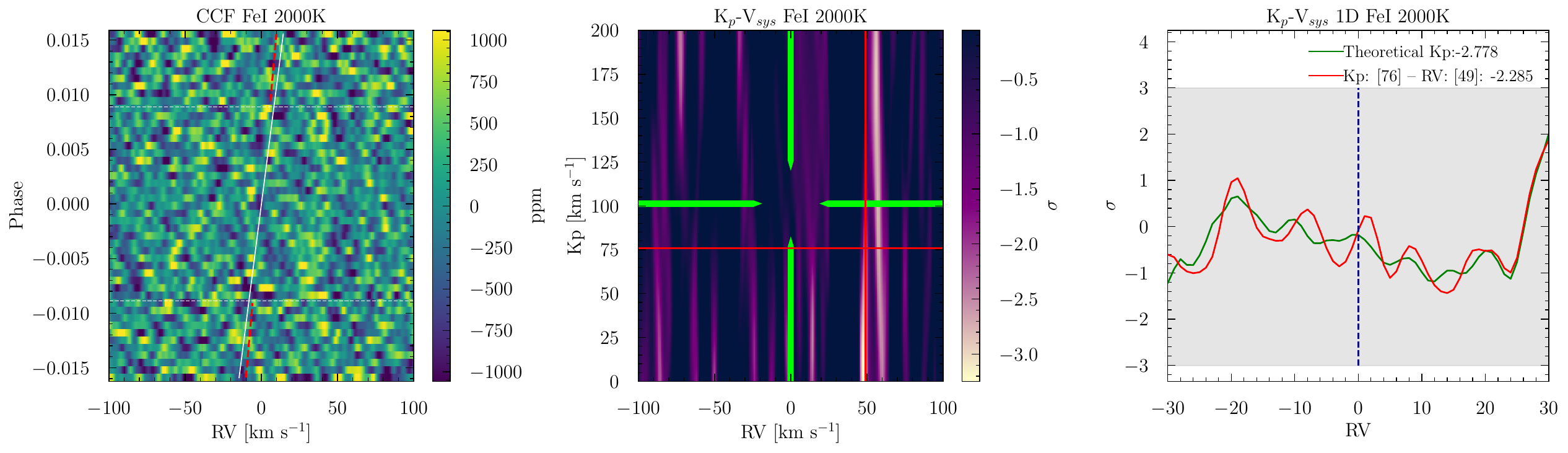}
    \caption{Cross-correlation and K$_p$-V$_{sys}$ maps of Na {\sc i}, Ca {\sc i,} and Fe {\sc i} obtained with the \cite{2023A&A...669A.113K} templates at 2000K. In the first column the CCF are depicted in terms of ppm. The dotted horizontal lines indicate the T1 and T4 contact points, while the red dashed line is the planetary trace. The Doppler shadow follows the white slanted line. Second columns are the K$_p$-V$_{sys}$ maps. The planetary signal is expected to be limited between the green pointer in the middle of the figures, while the red lines point at the minimum value of the K$_p$-V$_{sys}$ between K$_p$ 75 \kms, 125 \kms and RV -50 \kms, +50 \kms. The last column is the K$_p$-V$_{sys}$ 1D map evaluated at the K$_p$ minimum value. The grey region is the 3$\sigma$ edge, while the vertical navy dashed line refers to the RV 0 \kms.}
    \label{fig:ccf_2000K}
\end{figure*}

\clearpage
\section{Comparison with SLOPpy}
\label{app:sloppy}
\begin{figure*}[h]
                                \includegraphics[width=0.47\textwidth]{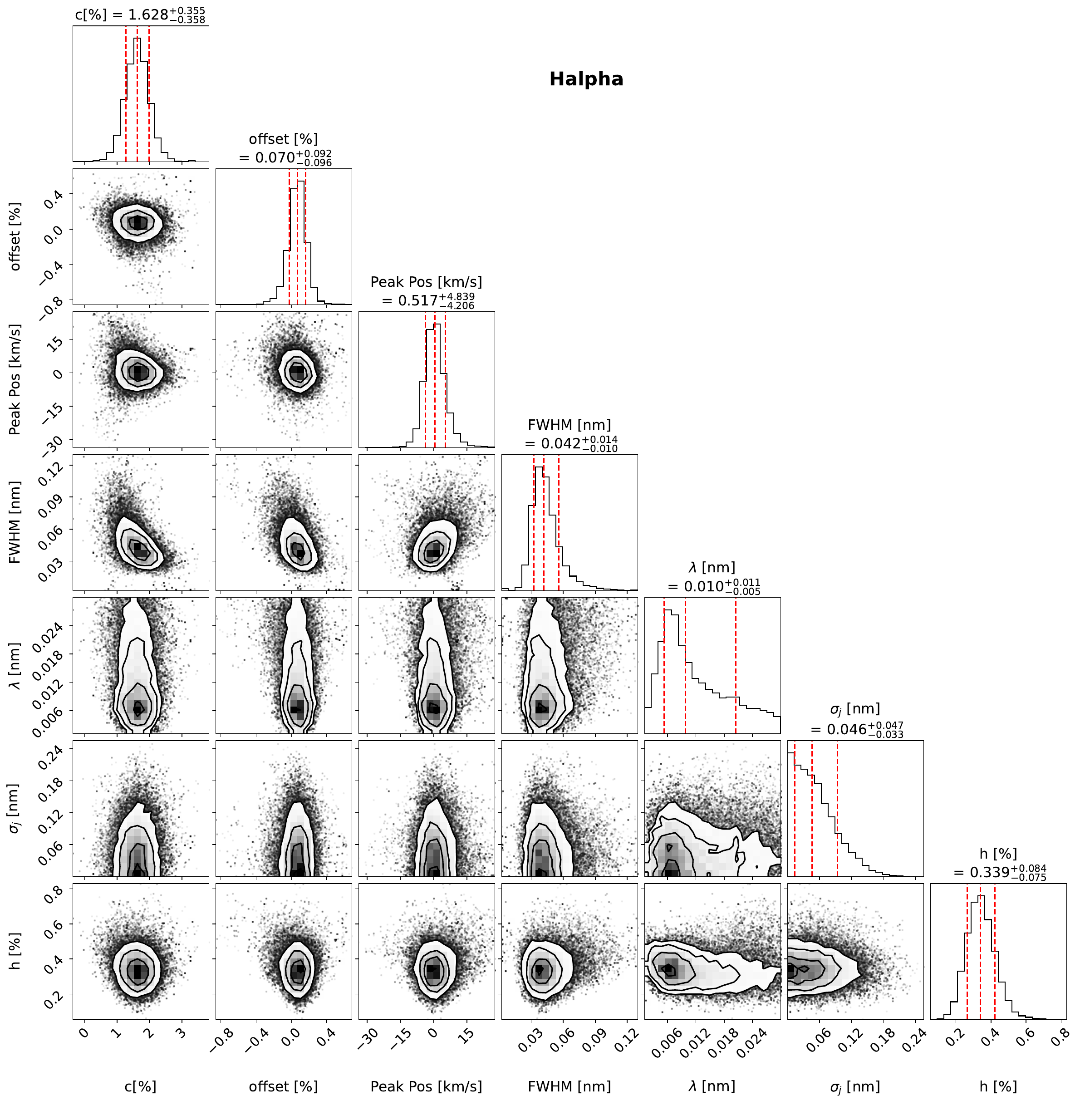}\\
                                \includegraphics[width=0.47\textwidth]{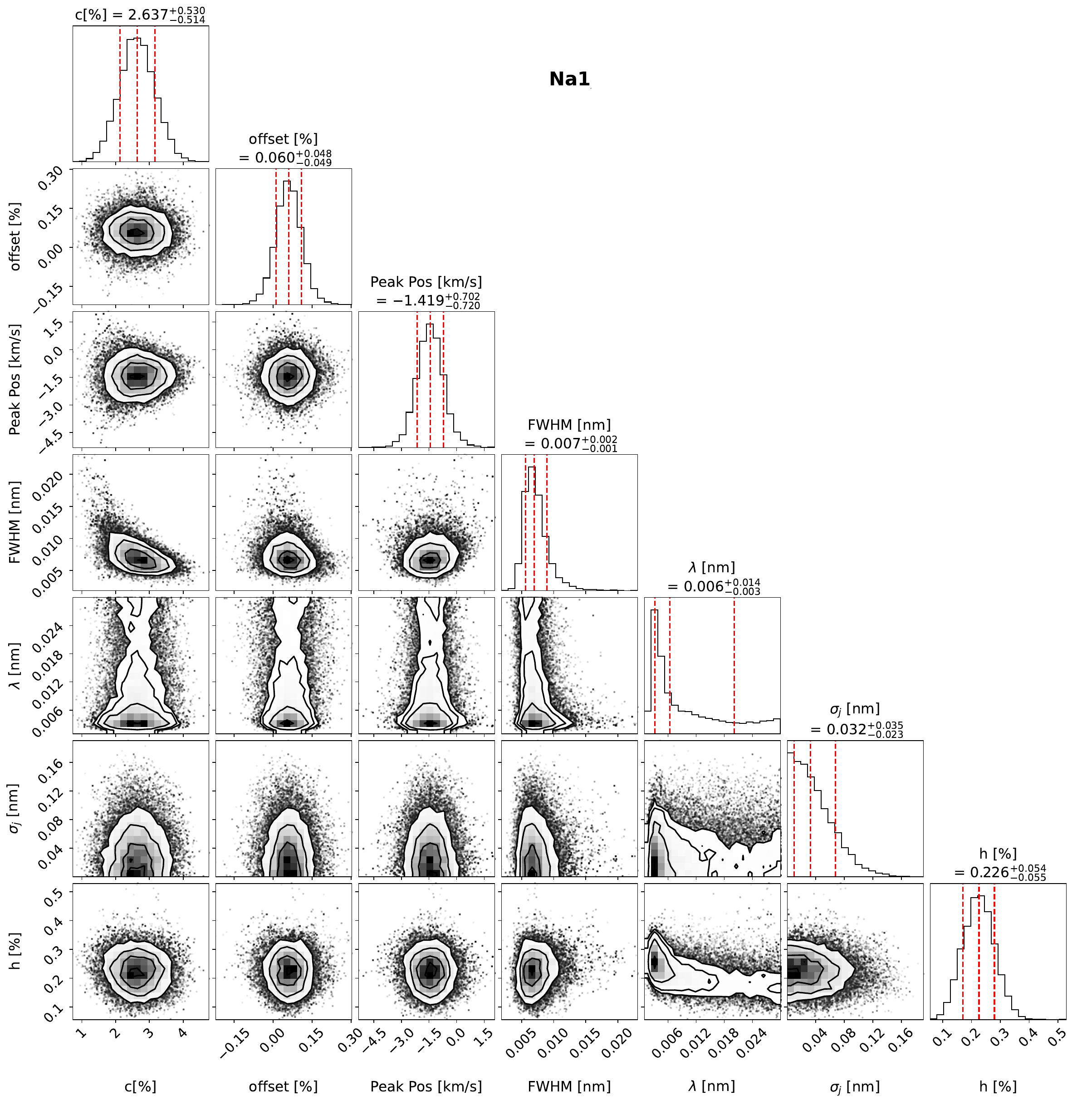}
                                \includegraphics[width=0.47\textwidth]{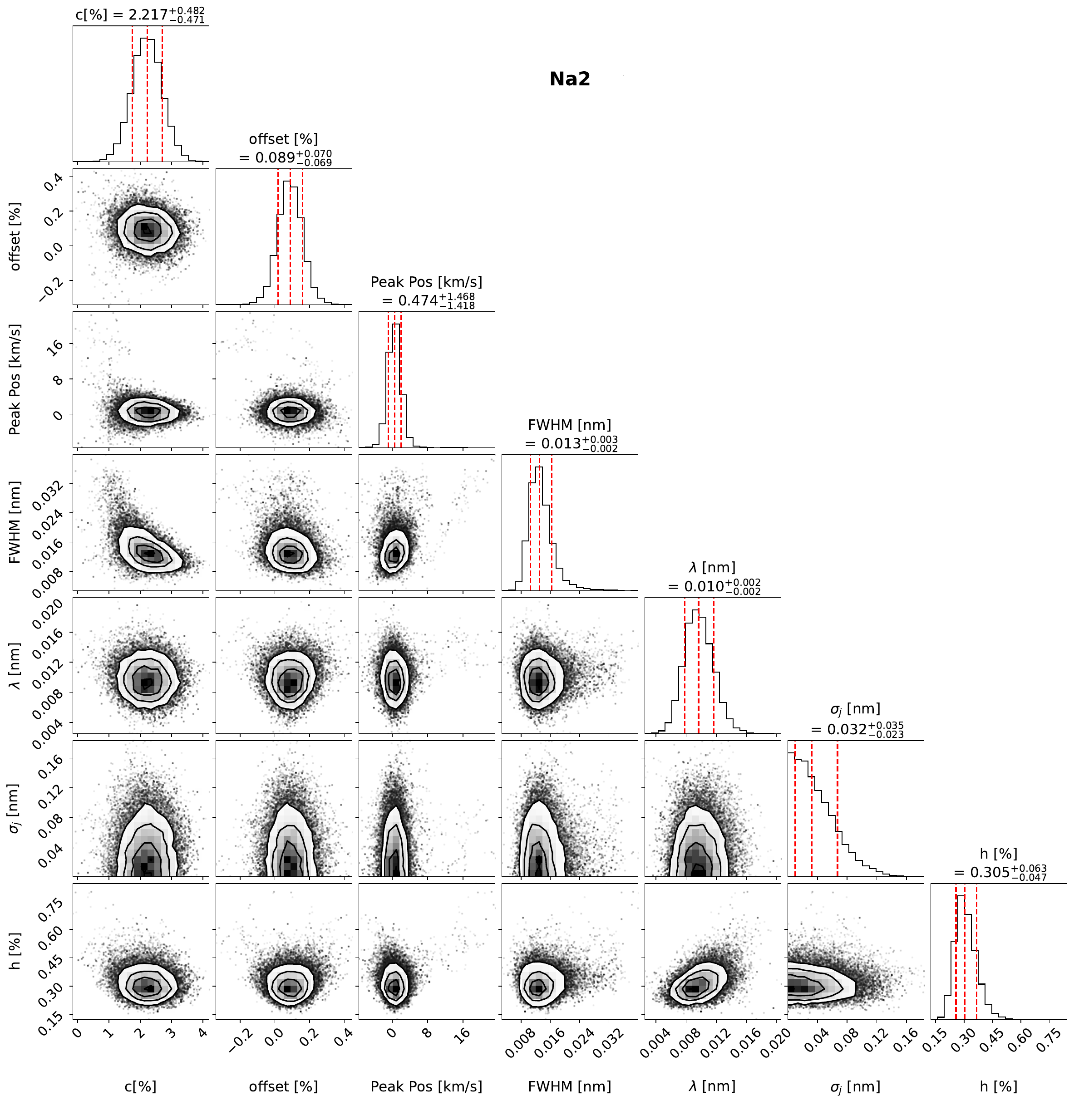}
                                \label{corner_sloppy}
        \caption{Same as Fig. \ref{Cornerplots} but for SLOPpy.}
        \vspace{1cm}
     \centering
     \captionof{table}{Best-fit parameters with SLOPpy.}
     \begin{tabular}{c | c | c | c | c | c | c}
                \hline \hline
                Line & \multicolumn{2}{|c|}{Peak position} & Contrast $c$ &  R$_\mathrm{eff}$ \tablefootmark{a} & FWHM & Significance\\
     &           [nm] & [\kms]     &         [\%]   & [Rp]      & [nm]   & [$\sigma$]        \\ 
    \hline
H$\alpha$ & 656.2837 $^{+ 0.0055 }_{ -0.0053 }$ &0 .52 $^{+ 4.84 }_{ -4.21}$ & 1.63 $^{+ 0.35 }_{ -0.36 }$ & 0.92 $\pm$ 0.29 & 0.042 $^{+ 0.014 }_{ -0.010 }$ & 4.5  \\
NaD1 & 589.5901 $^{+ 0.0011 }_{ -0.0011 }$& -1.41 $^{+ 0.70 }_{ -0.72}$ & 2.64 $^{+ 0.53 }_{ -0.51 }$ & 1.12 $\pm$ 0.11 & 0.007 $^{+ 0.002 }_{ -0.001 }$ & 5.0 \\
NaD2 & 588.9953 $^{+ 0.0017 }_{ -0.0016 }$& 0.47 $^{+ 1.47 }_{ -1.41}$  & 2.21 $^{+ 0.48 }_{ -0.47 }$ & 1.12 $\pm$ 0.11 & 0.013 $^{+ 0.003 }_{ -0.002 }$  & 4.6 \\            
\hline
        \end{tabular}
        \tablefoot{Same as Table \ref{tab_result_combined} but for SLOPpy results. \tablefoottext{a}{The R$_\mathrm{eff}$ is fitted in the analysis by SLOPpy and it is not retrieved from the contrast. Therefore we did not include the $\delta_\mathrm{R_P}$/H$_{\mathrm{eq}}$ as in Table \ref{tab_result_combined}}.}
        \label{tab_sloppy}
\end{figure*}

We compared the results obtained in the visible range with the method described in this work with SLOPpy to evaluate the impact of different correction of the RME+CLV. The two methods differ by many factors but rely on the same idea of correcting for the part of the stellar disc obscured by the planet. Our method sets the extension of the atmosphere increasing iteratively the radius retrieved from the Gaussian fit until we reach the convergence of 0.001 Rp, as described in Sect. \ref{vis_method} and hence we retrieve the $R_\mathrm{eff}$ from the absorption depth. On the other hand, SLOPpy fits the size of the atmosphere by comparing the 2D map containing the modelled RME+CLV and the planetary signal employing an MCMC approach. We stress that the extension retrieved by SLOPpy is not the physical extension of the atmosphere but a corrective factor.
Therefore, the major difference in the two frameworks is the extension adopted to perform the RME + CLV correction. We performed the comparison in the following way: we employed SLOPpy which returns the extracted transmission spectrum and fit it using an MCMC method to find the best Gaussian fit values and the best corrective radius factor to employ in the RME+CLV correction. Then, we applied the same GP+DEMCMC analysis described in Sect. \ref{sec:methods} on the spectrum extracted with SLOPpy. The results are shown in Fig. \ref{corner_sloppy} and Table \ref{tab_sloppy}. The FWHM and the velocity shift are within 1$\sigma$ for both the H$\alpha$ and Na {\sc i} doublet, while, for the latter, the contrasts we retrieve are larger than SLOPpy's ones. A deep comparison between the two frameworks is out of the scope of this work and hence we limit to say that, despite the difference in depth, in both cases we detect the presence of the two Na lines anf the H$\alpha$ in the atmosphere of planet b. 

\end{appendix}
\end{document}